\documentclass[journal=jacsat,manuscript=article]{achemso}

\usepackage{amsmath,amssymb}
\usepackage{xcolor}

\author{Shay Blum}
\affiliation{Schulich Faculty of Chemistry and Helen Diller Quantum Center, Technion-Israel Institute of Technology, Haifa 3200003, Israel}
\alsoaffiliation{Physics Department, Technion-Israel Institute of Technology, Haifa, 3200003, Israel}
\affiliation{Schulich Faculty of Chemistry and Helen Diller Quantum Center, Technion-Israel Institute of Technology, Haifa 3200003, Israel}
\author{David Gelbwaser-Klimovsky}
\email{dgelbi@technion.ac.il}
\affiliation{Schulich Faculty of Chemistry and Helen Diller Quantum Center, Technion-Israel Institute of Technology, Haifa 3200003, Israel}

\title{Thermalization without detailed balance: population oscillations in the absence of coherences}

\begin{document}

\newpage

\begin{abstract}
Open quantum systems that comply with the master equation and detailed balance decay in a non-oscillatory manner to thermal equilibrium. Beyond the weak coupling limit, systems that break microreversibility (e.g., in the presence of magnetic fields) violate detailed balance but still thermalize. We study the thermalization of these systems and show that a temperature rise produces novel exceptional points that indicate a sharp transition in the thermalization dynamics. A further temperature increase fuels oscillations of the energy level populations even without quantum coherences. Moreover, the violation of detailed balance introduces an energy scale that characterizes the oscillatory regime at high temperatures.\end{abstract}

\begin{center}
\textbf{Table of Contents (TOC) Graphic}
\end{center}
\begin{figure}
	\centering
		\includegraphics{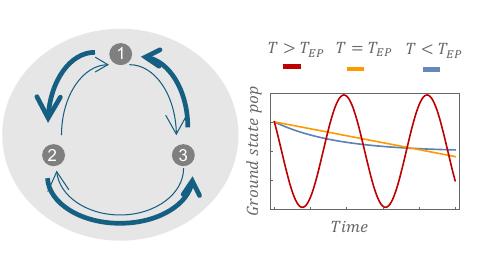}
	\label{fig:TOC}
\end{figure}

\maketitle
\newpage

Thermalization is the irreversible evolution towards a thermal equilibrium distribution. This process establishes a clear direction of time, even though the underlying dynamic equations are reversible in time.
The emergence of irreversibility and thermalization from a reversible evolution has been the source of multiple discussions \cite{boltzmann1872weitere,wehrl1978general,international1973boltzmann}. For quantum systems, thermalization was first studied by Pauli \cite{pauli1928sommerfeld}. 
Starting from the reversible Schr\"{o}dinger equation and assuming random phases, Pauli used perturbation theory to derive a rate equation for the energy level populations that explained how a quantum system in contact with its surroundings irreversibly reaches a thermal distribution. Since then, the Pauli master equation has been the standard equation for describing the thermalization of open quantum systems \cite{van1954quantum,isihara2013statistical}.

Detailed balance (DB) has been considered as a necessary condition for thermalization\cite{lewis_new_1925,fowler_note_1925}, hampering the study of thermalization beyond DB. The role of DB in thermalization becomes evident in the Pauli rate equation. In this equation, the constraints DB imposes on the transition rates force the system to evolve towards a thermal steady state through a non-oscillatory exponential decay of the energy levels' population. This is achieved by balancing the probability flow between each pair of eigenstates. Moreover, DB ensures the lack of persistent probability and heat currents at thermal equilibrium \cite{zia_probability_2007}.

Nevertheless, several systems can violate DB at equilibrium: photonic crystals \cite{zhu_near-complete_2014}, non-reciprocal planar slabs \cite{khandekar_new_2020} and electrons in quantum rings \cite{entin1995orbital}, to mention a few examples. As recently has been shown in the context of Lindblad and Pauli rate equations\cite{alicki2023violation}, when these systems interact only with a thermal bath, the steady state is a thermal equilibrium state despite the violation of detailed balance at equilibrium (VDB). Systems that violate DB present interesting features at thermal equilibrium such as persistent currents \cite{zhu_persistent_2016}, repulsive Casimir forces \cite{gelbwaser-klimovsky_equilibrium_2022} and potential violations of Earshaw's theorem \cite{fiedler2023perspectives}.  However, these studies focus on systems at thermal equilibrium, without investigating the thermalization process itself. In contrast, in the realm of algorithms, the VDB has been shown to accelerate the convergence of Markov Chain Monte Carlo methods \cite{ichiki2013violation}. This suggests the potential for VDB to influence the thermalization of real physical systems, a topic that remains unexplored. 
 
In this letter, we use the Lindblad equation to study the thermalization of a non-degenerate N-level open quantum system that violates DB. We show that the VDB enables different 
 types of dynamics that are absent for systems that keep DB, and therefore, they were unknown in the context of thermalization. The new thermalization dynamics are oscillatory decay and polynomial decay. The latter is produced by the presence of novel Liouvillian exceptional points (LEPs) \cite{minganti2019quantum} in the rate equations.
 We show that the resulting dynamics depends on the balance between the strength of two types of processes that the system undergoes simultaneously: relaxation and a tendency to oscillate in close loops along energy state populations. As we explain below, the latter process is favored at high temperatures and strong VDB. 

LEPs have been widely studied in settings with an external driving \cite{am2015exceptional,am2016parameter} or in the presence of multiple baths \cite{khandelwal2021signatures, minganti2019quantum} and have been shown to accelerate relaxation dynamics \cite{zhou2023accelerating}. Here, we find LEPs in a different scenario: an open quantum system without driving and interacting with a single thermal bath. To the best of our knowledge, this is the first work on LEPs effects on thermalization in an equilibrium setup at non-zero temperatures. The LEPs in our system are a consequence of the VDB and 
 they set a temperature, $T_{EP}$, where there is a sharp transition from a non-oscillatory to an oscillatory decay to the thermal state. 

 Although energy level oscillations can be found in several systems, 
 they are damped at high temperatures \cite{daniel2013damping} and require quantum coherences (in the eigenbasis of the system Hamiltonian) \cite{leggett1987dynamics}. The connection between population oscillations and quantum coherences has been considered as a signature of quantumness in a wide range of fields such as open quantum systems \cite{dodin2024population}, quantum thermodynamic \cite{mitchison_coherence-assisted_2015}, quantum biology \cite{engel2007evidence} and quantum transport \cite{timmer2023plasmon}.
 The oscillations we study are different for two fundamental reasons: i) They do not require quantum coherences between energy levels; ii) they arise only at $T>T_{EP}$ because they require the injection of energy that can only be provided by a high-temperature bath.

 Finally, we study the thermalization of a toy model composed of a single electron tunneling among three quantum dots in the presence of a magnetic field \cite{delgado_theory_2007}. This model allows us to derive analytically the required physical conditions for decaying oscillations. Moreover, in this model, we show that \emph{the VDB introduces an energy scale}, $\mathcal{E}_{VDB},$ that determines the regime of oscillatory behavior at high temperatures.

We  start by considering a non-degenerate N-level system interacting
with a \emph{single} thermal bath at inverse temperature $\beta=1/k_{B}T$. The reduced dynamics follows the Gorini, Kossakowski, Lindblad and
Sudarshan (GKLS) equation \cite{lindblad_generators_1976, gorini_completely_1976}. The latter is approximately valid in the weak coupling, low-density or singular coupling limit \cite{breuer_theory_2002}. We further assume that the N-level system is not nearly degenerate, therefore ensuring the accuracy of the global GKLS equation instead of the local one \cite{landi2022nonequilibrium}. The coherences and populations dynamic equations decoupled from each other and the Pauli rate equation describes the population evolution:

\begin{equation}
\mathbf{\dot{P}=}M\mathbf{P},\label{eq:dyn}
\end{equation}
where $\mathbf{P}$ is a vector composed of the populations of the
system energy levels and $M$ is the transition rate matrix with components
$M_{k\neq l}=a_{kl}$ and $M_{kk}=-\sum_{l\neq k }a_{lk}$. $a_{kl}$
represents the transition rate from state $l$ to $k$ and 
it is derived from microscopic dynamics.
For systems that are weakly coupled or keep microreversibility,
$a_{kl}$ obeys DB \cite{alicki_quantum_2007}, that is, $a_{kl}e^{-\beta\mathcal{E}_{l}}=a_{lk}e^{-\beta\mathcal{E}_{k}}.$
Here $\mathcal{E}_{k}$ is the energy of the system's level $k$. Beyond the weak coupling limit, systems that do not keep microreversibility (e.g., systems in the presence of magnetic fields) may violate DB, but they still relax to a thermal state
\cite{alicki2023violation}. Their transition rates comply with more complex constraints
known as thermalization conditions or complex balancing \cite{feinberg2019foundations}.

Despite the stationary state being the same for transition rates that keep or violate DB, \emph{how the thermal state is reached is fundamentally different}. Systems that violate DB simultaneously experience two types of processes: $i)$ Standard dissipation,
which relaxes the system state towards the thermal state,
reducing the relative entropy between these two states. The strength of this process is characterized by $\omega_{dis}=\sum_{k\neq l}a_{kl}>1/t_{dis}$, where $t_{dis}$ is the thermalization time scale;
$ii)$ A tendency to oscillate along closed loops among the population
states, for example: $|l\rangle \rightarrow |k\rangle \rightarrow |i \rangle\rightarrow...\rightarrow | j\rangle \rightarrow |l\rangle.$ We emphasize that these are oscillations of the system Hamiltonian eigenstates and \emph{do not require the presence of quantum coherences} among these levels. These oscillations are generally absent 
in scenarios without driving or multiple thermal baths because they require the VDB. This is demonstrated by the violation of Kolmogorov’s criterion \cite{kelly2011reversibility}: $c=a_{lj}...a_{ik}a_{kl}-a_{lk}a_{ki}...a_{jl}\neq0$, indicating a difference between the forward and backward processes, and implying a non-zero affinity \cite{biddle2020reversal}. On multilevel systems there could be several different closed loops among population states. DB implies $c=0$ for all possible closed loops and therefore,
no oscillations. As we show below, the balance between the strengths of the two types of processes ($|c|$ vs $\omega_{dis}$) determines the thermalization dynamics. 

The general solution of Eq. \eqref{eq:dyn} is 

\begin{equation}
\mathbf{P
}(t)=\mathbf{P}_{th}+\sum_{i=1}^{N-1}\sum_{l=0}^{d_{i}-1}b_{i,l}t^{l}e^{\lambda_{i}t}\mathbf{V}_{i},
\label{eq:pauli_sol}
\end{equation}
where $\mathbf{P}_{th}$ is the thermal distribution and $b_{i,l}$ are parameters set by the initial conditions. $\lambda_{i}$ and $\mathbf{V}_{i}$ are eigenvalues and generalized eigenvectors of $M$, respectively. $d_{i}$ is the number of generalized eigenvectors related to the eigenvalue $\lambda_{i}$ (see Sec. S.1 on the Supplemental Material). 
If the transition rates keep DB, then all $\lambda_{i}$
are real and non-positive, and $d_i=1$. The system thermalizes through a linear combination of decaying exponentials. We term this dynamics non-oscillatory decay. If the transition rates violate DB,
the thermalization dynamics can be different. Besides the non-oscillatory decay, there
are two other dynamical regimes: 1) For large enough $c$ and temperature, some of
the eigenvalues become complex with a non-positive real part, and
the dynamics acquires an oscillatory decay component. Here too $d_i=1$. Although these oscillations may be used for a transient work extraction,   it is based on the resources present in the initial state. The lack of steady work extraction ensures compliance with the second law of thermodynamics;
2) The regime between oscillatory and non-oscillatory decay is divided
by the exceptional point dynamics. In this case, $M$ has some real non-positive eigenvalues that are
degenerate and their respective eigenvectors coalesce. $M$ is no
longer diagonalizable and its Jordan form is required for the derivation of \eqref{eq:pauli_sol} ($d_i>1$ for at least one $i$). The relaxation to the thermal state acquires a polynomial component that multiplies the exponential decay. Exceptional points of $M$ are termed LEPs \cite{minganti2019quantum} because they are related to the Liouvillian quantum dynamics of open systems, rather than the Hamiltonian dynamics. Reaching all the dynamical regimes is not ensured by the VDB. Physical transition rates are limited by the thermalization conditions. As shown below, even under those constraints, quantum systems can experience LEP and oscillatory thermalization dynamics.  

Temperature and  the strength of the VDB  determine the thermalization dynamics. Because the system is non-degenerate, oscillations in close loops along the populations need energy that can only be provided by the bath. If the temperature is low, the bath can not supply the required energy and the system relaxes exponentially to equilibrium with no oscillations (see Sec. S.6 on the Supplemental Material). For higher temperatures, the bath can 
fuel the oscillations. In this case, if the VDB is strong enough (i.e., large $|c|$),
the thermalization dynamics will exhibit an oscillatory decay. To determine the required amount of the VDB for triggering oscillations, we study the simplest open system that can violate DB without driving or multiple baths: an open three-level system toy model.

 \begin{figure}
 \centering
 \includegraphics[width=0.75\textwidth]{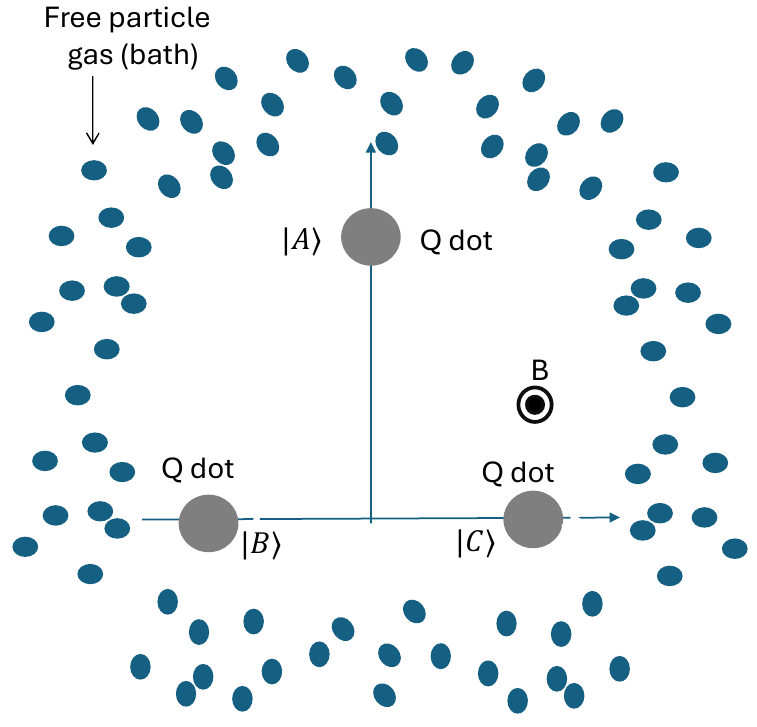}
 \caption{Toy model. A free particle gas (thermal bath) is scattered by a single electron that tunnels among three QDs in a magnetic field. In this work, we assume a short separation between the three quantum dots and a 1D gas. These assumptions allow us to calculate analytically the $\mathcal{T}$-matrix elements.}
 \label{fig:sketch}
\end{figure}

For this, we consider three quantum dots (3QDs)
in an equilateral triangle arrangement under the influence of a magnetic
field \cite{delgado_theory_2007} (see Fig. \ref{fig:sketch}). The QDs positions are $\mathbf{r}_{i}$. Assuming there is only
a single electron in the system, its Hamiltonian in the single electron
localized basis ($|1\rangle,|2\rangle,|3\rangle)$ is 
\begin{equation}
H_{el}=\tau\left(\begin{array}{ccc}
0 & e^{-i2\pi\phi/3} & e^{i2\pi\phi/3}\\
e^{i2\pi\phi/3} & 0 & e^{-i2\pi\phi/3}\\
e^{-i2\pi\phi/3} & e^{i2\pi\phi/3} & 0
\end{array}\right), \label{eq:ham} 
\end{equation}
here $\tau$ is the tunneling constant and $\phi$ is the magnetic
flux quanta, which allows the breaking of microreversibility. The
eigenenergies of this Hamiltonian are $\mathcal{E}_{+}>\mathcal{E}_{0}>\mathcal{E}_{-}.$
The 3QDs interact with a low-density gas of free particles with mass
$m$. The gas is in a thermal state with inverse temperature $\beta=1/k_B T$. We assume
that if there is an electron in the quantum dot, a nearby particle
will feel a short-range repulsive potential. In particular, we model the interaction Hamiltonian
 as $H_{int}=\sum_{i\in \{1,2,3\}}V_{i}\delta(\mathbf{r}-\mathbf{r_{i}})|i\rangle\langle i|$. Microrevesibility and DB are recovered if $V_i=V_j$ for any pair with $i\neq j$. 
 The low density of the gas allows to describe the 3QDs reduced dynamics
with the low-density limit GKLS equation. At this limit, the gas statistic does not play any role\cite{dumcke_low_1985}. Here, the role of the jump operators is fulfilled by the on shell $\mathcal{T}$-matrix elements describing the scattering of the low-density gas particles by the electron of the 3QDs. $|\langle \textbf{p}'k|\mathcal{T}|\textbf{p}l\rangle|^{2}$ represent the probability
for a process that starts with the 3QDs in a state $l$ and a gas particle with momenta $\textbf{p}$ and ends with the 3QDs at state $k$ and the scattered gas particle with momenta $\textbf{p}'$. For a short separation between the three quantum dots and a 1D gas, the $\mathcal{T}$-matrix elements
can be calculated analytically, and from them, the transition rates are obtained by tracing out the incoming and outgoing gas particle momenta (see Sec. 4 on the Supplemental Material):

\begin{gather} 
 a_{kl}=2\nu\pi\intop dp\,dp'\,e^{- \frac{\beta p^2}{2m }}Z_{P}^{-1}\delta\left[\frac{p'^2}{2m}+\mathcal{E}_{k}-\left(\frac{p^2}{2m}+\mathcal{E}_{l}\right)\right] \left|\left\langle k,p'\left|\mathcal{T}\right|l,p\right\rangle \right|^{2}, \label{eq:rates}
\end{gather}
where $Z_{P}$ is the partition function of the gas particle and $\nu$ is the gas density. These transition rates derived from a physical microscopic model are used to build the transition matrix $M$.

\begin{figure}
 \centering
 \includegraphics[width=0.75\textwidth]{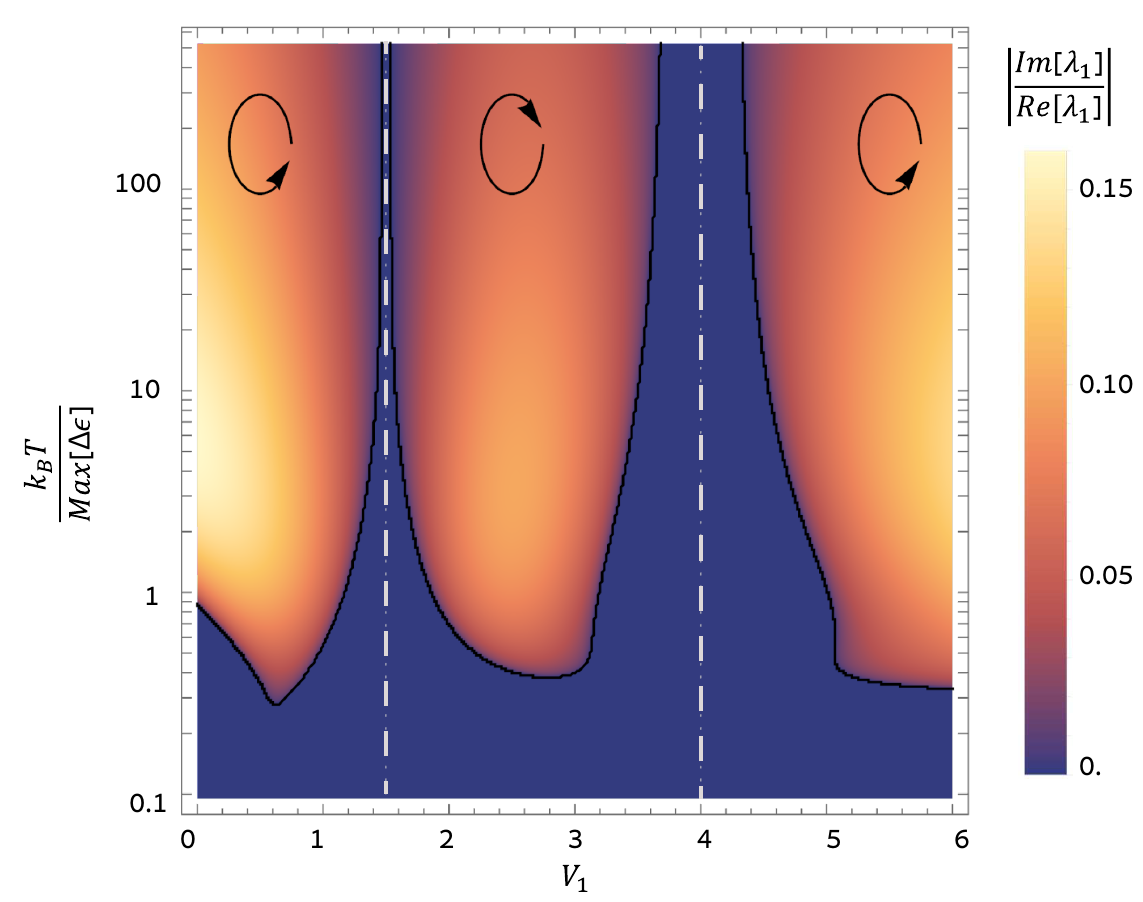}
 \caption{3QDs thermalization regimes as function of $V_1$ (x-axis) and temperature $T$ (y-axis). $V_1$ controls the VDB. The plot shows the absolute value of the ratio between the imaginary and real part of the non-zero eigenvalue of $M$. For non-oscillatory decay (blue regions) eigenvalues are real and the plotted rate is zero. For oscillatory decay (yellow/orange regions) eigenvalues can be complex and the plotted rate gets a non-zero value. Black lines separating oscillatory and non-oscillatory decay correspond to the exceptional point dynamics and determine $T_{EP}$. Gray dashed vertical lines correspond to regions where microreversibility and DB are reestablished ($V_1=V_2=1.5$ and $V_1=V_3=4$). Parameters: $\tau=1.85$, $\phi=0.575$ and $\hbar=k_B=m=1$. Max $\Delta \epsilon = \mathcal{E}_+-\mathcal{E}_-$. } 
 \label{fig:regimes}
\end{figure}

Figure \ref{fig:regimes} shows the number of oscillations during the system thermalization time scale as a function of the temperature $T$ and $V_1$. $V_1$ determines the strength of the VDB and the asymmetry of the dynamics, which sets the oscillation direction. $T$ fixes the bath energy that drives the oscillations. The number of oscillations indicates the different dynamical regimes and is determined by the ratio between the imaginary and real parts of $M$ non-zero eigenvalue.
 As discussed above, the system always decays without oscillations to the thermal state for low temperatures. This changes
at higher temperatures. If the VDB 
is large enough, there is a temperature, $T_{EP}$, at which a LEP is formed and a sharp transition on the thermalization dynamics takes place. 
Above $T_{EP}$ the system will thermalize through decaying oscillations. Notice that the regions that comply with DB (i.e., $V_{1}= V_{2}$ or $V_{1}= V_{3}$) divide between
oscillatory decay with different \emph{directions}: clockwise in the middle region $\left(|-\rangle\rightarrow |0\rangle \rightarrow |+\rangle \rightarrow |-\rangle\right)$ where the rates in the clockwise direction, $\circlearrowright$, ($a_{0-}, a_{+0}$ and $a_{-+}$) are the largest rates; and counterclockwise, $\circlearrowleft$, in the lateral regions $\left(|-\rangle\rightarrow |+\rangle \rightarrow |0\rangle \rightarrow |-\rangle \right)$ where the counterclockwise direction rates are the dominant rates ($a_{+-}, a_{0+}$ and $a_{-0}$). The plotted oscillations number never goes above $0.16$. However, the oscillation number can be increased by considering larger systems \cite{tarnowski2021random} and increasing the strength of the VDB. It can reach 1 for a 4-level system and surpass 2 for a 7-level system (see Supplementary material section S.7).

The VDB or oscillation strength required to trigger
the oscillatory decay can be found by analyzing the
eigenvalues of $M$. The oscillation strength is the difference between the rate of the counterclockwise and clockwise   process, i.e.,
$c=a_{-+}a_{+0}a_{0-}-a_{-0}a_{0+}a_{+-}$. 
To have oscillatory decay, $c$ should keep the inequality (see Sec. S.2 on the Supplemental Material)

\begin{equation}
 \left|c\right|>\frac{\omega_{dis}^{2}\left|\omega_{dis,DB}-\omega_{dis}\right|}{4\sum_{k,l}e^{\beta\left(\mathcal{E}_{k}-\mathcal{E}_{l}\right)}},
 \label{eq:balance}
\end{equation}
where $\omega_{dis,DB}=\sum_{\{k,l\}\in \circlearrowright}a_{kl}\left(1+e^{\beta (\mathcal{E}_{k}-\mathcal{E}_{l})}\right)$ is the sum of the respective rates when DB holds. Eq. \eqref{eq:balance} confirms the physical intuition that the balance between oscillations and dissipation strengths, $|c|$ and $\omega_{dis}$ respectively, determines the thermalization dynamics. Eq. \eqref{eq:balance} becomes an equality at the LEP.

 The mechanism driving the system to an oscillatory decay can be understood
by analyzing the rates and the corresponding $\mathcal{T}$ matrix. For our system, the rates, \eqref{eq:rates}, can be rewritten in the following form, which simplifies the analysis (see Sec. S.4 on the Supplemental Material):

 \begin{gather} a_{kl}=\sqrt{\beta}e^{\beta\mathcal{E}_{l}}\int_{\mathcal{E}_{+}}^{\infty}dEe^{-\beta E}\left(a_{0}+\left(-1\right)^{q}a_{1}+b_{kl}\right)+\sqrt{\beta}e^{\beta\mathcal{E}_{l}}\int_{\mathcal{E}_{0}}^{\mathcal{E}_{+}}dEe^{-\beta E}\tilde{b}_{kl},
\label{eq:proc}
 \end{gather}
where $q=0$ for clockwise rates and $q=1$ otherwise.
 The terms $a_{0},a_{1},b_{kl}$ and $\tilde{b}_{kl}$ are independent of the temperature.
Oscillatory decay takes place at high temperatures. Raising the temperature increases the number of gas particles with high energy, which eventually can provide the dominant contribution to
$a_{kl}$. Therefore, the mechanism behind the oscillatory dynamics
can be understood from the high energy expansion of $a_{0}, a_{1}$ and $b_{kl}$ (see Sec. S.4 on the Supplemental Material):
\begin{equation}
 \left\{ a_{0},a_{1},b_{kl}\right\} \underset{E\gg\mathcal{E}_{+}}{\propto}\left|v_{k \neq l}\right|^2\left\{ \frac{1}{E},\,\frac{\sqrt{\mathcal{E}_{VDB}}}{E^{3/2}},\,\frac{\mathcal{E}_{j\ne\,(k,l)}}{E^{2}}\right\}, 
 \label{eq:lim}
\end{equation}
where $\mathcal{E}_{VDB}$ is the energy scale related to the VDB. In terms of the Hamiltonian parameters, $\mathcal{E}_{VDB}=\left(\frac{2\sqrt{6m}\Im[v_{-0}v_{0+}v_{+-}]}{\hbar|v_{k\neq l}|^2}\right)^2$. Here $v_{kl}=\sum_{i=1}^{3}\langle k|i\rangle V_{i}\langle i|l\rangle$,
$k,l\in\{+,0,-\}$, are the matrix elements
of the interaction Hamiltonian part that acts on the system. They have units of $Energy\times Length$.
For our system
$|v_{kl}|^{2}$ is the same for any $k\neq l$ and therefore $\mathcal{E}_{VDB}$ is well defined. Eq \eqref{eq:lim} shows that at high energies, the main contribution to the rates is $a_0$. $a_1$ and $b_{kl}$ are first and second-order corrections, respectively. 
For $E\gg\mathcal{E}_{+}$,
$a_0$ do not distinguish between transitions
unless the interaction Hamiltonian has some asymmetry such that $|v_{kl}|^{2}\neq|v_{lm}|^{2}$ ($k\neq l\neq m)$.
This makes $a_{0}$ transition independent for our model. The high energy term of
$a_{0}$ originates from the Born or weak coupling approximation of
the $\mathcal{T}$-matrix, which does not contribute to VDB \cite{alicki2023violation}. The VDB only arises 
at the next order of the $\mathcal{T}$-matrix Born series which is proportional
to {$\Im\left(v_{-0}v_{0+}v_{+-}\right)\propto v_{-+}v_{+0}v_{0-}- v_{-0}v_{0+}v_{+-} \propto \left(V_{1}-V_{2}\right)\left(V_{2}-V_{3}\right)\left(V_{1}-V_{3}\right)$ (see Sec. S.3 on the Supplemental Material). This quantity is the Hamiltonian equivalent to the difference between the clockwise and counterclockwise process
 and is proportional to the violation of microreversibility. This $\mathcal{T}$-matrix term produces the high energy limit of $\left(-1\right)^q a_{1}$,
 which does not distinguish among all the individual transitions
but makes a difference between clockwise and counterclockwise rates through the prefactor
$\left(-1\right)^q$. Next, there is
 $b_{kl}$ which is the first term in the $1/E$ series expansion
 that distinguishes between transitions in the same direction, but not between directions, i.e., $b_{kl}=b_{lk}$. Finally, $\tilde{b}_{kl}$
integral is limited to $\mathcal{E}_{+}$, so it does not include high energy
contributions.

 If the VDB is large enough, see Eq \eqref{eq:balance}, 
at high temperatures the high energy contributions of $a_1$ will overshadow
the low energy contributions of $b_{kl}$ and $\tilde{b}_{kl}$. In
this case, the rates in the same direction have approximately the
same value $a_{-+}\sim a_{+0}\sim a_{0-}\sim \sqrt{\beta}\int_{\mathcal{E}_{+}}^{\infty}dEe^{-\beta E}\left(a_{0}+a_{1}\right)$
 and $a_{-0}\sim a_{0+}\sim a_{+-}\sim\sqrt{\beta}\int_{\mathcal{E}_{+}}^{\infty}dEe^{-\beta E}\left(a_{0}-a_{1}\right)$. Under these circumstances, thermalization occurs through decaying oscillations with frequency proportional to $2\sqrt{\beta}\int_{\mathcal{E}_{+}}^{\infty}dEe^{-\beta E}a_{1}\propto \sqrt{\beta \mathcal{E}_{VDB}}$ (see Fig. \ref{fig:freq}). For weak VDB, the rates at high temperatures do not group into two different values depending on their direction and thermalization takes place through non-oscillatory relaxation (regions around $V_1\sim V_2=1.5$ and $V_1\sim V_3=4$ in figure \ref{fig:regimes}). 

\begin{figure}
 \centering
 \includegraphics[width=0.75\textwidth]{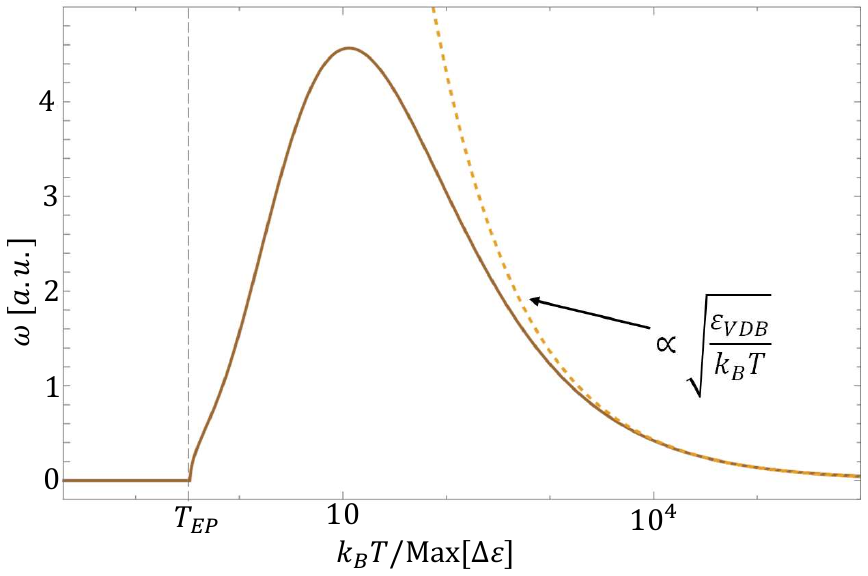}
 \caption{Populations oscillation frequency during thermalization (continuous, brown line) as a function of the bath temperature for a strong VDB (see Eq. \eqref{eq:balance}). At low temperatures, the system thermalizes through a non-oscillatory decay. At $T=T_{EP}$, there is a LEP, which produces a sharp transition to a thermalization through decaying oscillations. At high temperatures, the oscillation frequency is proportional to $\sqrt{\frac{\mathcal{E}_{VDB}}{k_BT}}$ (dashed yellow line), where $\mathcal{E}_{VDB}$ is the VDB energy scale (see Eq. \eqref{eq:lim}). Parameters: The same as figure \ref{fig:regimes} with $V_1=6$.}
 \label{fig:freq}
\end{figure}

Finally, we derive a \emph{sufficient} but not necessary condition for oscillatory decay at the high-temperature limit, i.e., $\beta \mathcal{E}_+ \ll 1$. For this, we introduce an energy scale $\mathcal{E}_{low-energy}$, which is related to low-energy processes (i.e., $\mathcal{E}_0\leq E\leq \mathcal{E}_+$ which are related to $\tilde{b}_{kl}$ and processes of order $\mathcal{O}\left(\frac{\mathcal{E}_k-\mathcal{E}_l}{E^3}\right)$, see Eq. \eqref{eq:lim}). At high temperatures, if

\begin{gather}
 \mathcal{E}_{VDB}>\frac{\left(\mathcal{E}_--\mathcal{E}_0\right)^2+\left(\mathcal{E}_--\mathcal{E}_+\right)\left(\mathcal{E}_0-\mathcal{E}_+\right)}{\mathcal{E}_+}+\mathcal{E}_{low- energy}, \label{eq:bound}
\end{gather}
the thermalization dynamics has to be through decaying oscillations. This inequality was derived using the high energy expansion of the rate processes at the high-temperature limit (see Sec. S.5 on the Supplemental Material). For lower temperatures, Boltzmann factors in the rates can not be neglected, complicating the derivation of a compact energetic condition as Eq. \eqref{eq:bound}.

In summary, the VDB produces alternative thermalization paths that result in two phenomena: i) the existence of LEPs at equilibrium conditions (i.e., without driving and in the presence of a single thermal bath). These novel LEPs produce a sharp transition in the thermalization dynamics, triggering oscillations. These LEPs could be used to expand the EPs' advantages for sensing \cite{wiersig2014enhancing,wiersig2020review} to 
thermal equilibrium settings, allowing the creation of more precise measurement protocols of equilibrium variables such as temperature. This will require to overcome the impact of noise \cite{lau_fundamental_2018,loughlin2024exceptional}; ii) Unique energy level population \emph{oscillations without quantum coherence} that instead of being damped they are fueled by high-temperature thermal noise. At high temperatures, the frequency of these oscillations, $\omega$, is determined by the VDB natural energy scale, $\mathcal{E}_{VDB}$: $\omega \propto \sqrt{\mathcal{E}_{VDB}/(k_BT)}$. Moreover, the relative value of $\mathcal{E}_{VDB}$ to other energy scales sets conditions for oscillations.

\textbf{Supporting information} The supporting information add details on: 1) the General solution to the Pauli master equation; 2) Eigen-decomposition of the transition rate matrix and condition
for oscillations; 3) Derivation of rates of transitions and the T matrix elements; 4) Decomposition of the rates of transition; 5) Sufficient condition for oscillations at the high-temperature
limit; 6) Proof for no oscillations in the low-temperature limit; 7) N - level system toy model.

\textbf{Acknowledgement}
We thank Milan \v{S}indelka, Ohad Cremerman, Thales Pinto Silva, Uri Peskin, Karol Życzkowski, Sergiy Denysov, Dariusz Chruscinski and Natan Granit for valuable discussions. D.G.K. is supported by the ISRAEL SCIENCE FOUNDATION (grant No. 2247/22) and by the Council for Higher Education Support Program for Hiring Outstanding Faculty Members in Quantum Science and Technology in Research Universities.

\providecommand{\latin}[1]{#1}
\makeatletter
\providecommand{\doi}
  {\begingroup\let\do\@makeother\dospecials
  \catcode`\{=1 \catcode`\}=2 \doi@aux}
\providecommand{\doi@aux}[1]{\endgroup\texttt{#1}}
\makeatother
\providecommand*\mcitethebibliography{\thebibliography}
\csname @ifundefined\endcsname{endmcitethebibliography}  {\let\endmcitethebibliography\endthebibliography}{}

\newpage
\newpage

\setcounter{equation}{0}
\renewcommand{\theequation}{S\arabic{equation}}

\setcounter{figure}{0}
\renewcommand{\thefigure}{S\arabic{figure}}

\setcounter{section}{0}
\renewcommand{\thesection}{S\arabic{section}}

\setcounter{subsection}{0}
\renewcommand{\thesubsection}{\thesection.\arabic{subsection}}

\begin{centering}
{\large \bf Thermalization  without detailed balance:  population oscillations in the absence of coherences: Supplementary information}\\
\end{centering}

\section{General solution to the Pauli master equation}
The solution for equation (1) in the main text is given by: $\mathbf{P}\left(t\right)=e^{Mt}\mathbf{P}_{0}$ where $\mathbf{P}_{0}=\mathbf{P}\left(t=0\right)$ sets the initial conditions. In order to arrive at equation (2) in the main text, one should consider the Jordan canonical form $J$ of the matrix $M$. Generally it has the form $J=diag\left\{ J_{1},J_{2},...,J_{\kappa}\right\} $ where $J_i$ are Jordan blocks related to eigenvalue $\lambda_i$, with dimension $d_i$. Note that by definition $\sum_{i=1}^\kappa d_i=N$, where $N$ is the dimension of $M$.\\
Considering the basis transfer matrix $Q$, we have:
\begin{equation}
    \label{eqn:dynamics_sol}\mathbf{P}\left(t\right)=e^{Mt}\mathbf{P}_{0}=e^{QJQ^{-1}t}\mathbf{P}_{0}=Qe^{Jt}Q^{-1}\mathbf{P}_{0}
\end{equation}
From the structure of $J$,  we have $e^{Jt}=diag\left\{ e^{J_{1}t},e^{J_{2}t},...,e^{J_{\kappa}t}\right\}$, where the exponential of a Jordan block is 
\begin{equation}    \left(e^{J_{i}t}\right)_{k\ell}=e^{\lambda_{i}t}\begin{cases}
\frac{1}{\left(\ell-k\right)!}t^{\ell-k} & k\le\ell\\
0 & k>\ell
\end{cases}
\end{equation}
Denoting $\tilde{\mathbf{P}}\left(t\right)=Q^{-1}\mathbf{P}\left(t\right)$, we can define the vectors $\mathbf{\tilde{P}}_{0}^{n}$ by
\begin{equation}
    \left(\mathbf{\tilde{P}}_{0}^{n}\right)_{i}=\begin{cases}
\left(\mathbf{\tilde{P}}_{0}\right)_{i} & \sum_{s=1}^{n-1}d_{s}<i\le\sum_{s=1}^{n}d_{s}\\
0 & \text{else}
\end{cases}
\end{equation}
such that $\mathbf{\tilde{P}}_{0}=\sum_{n=1}^{\kappa}\mathbf{\tilde{P}}_{0}^{n}$ and 
\begin{equation}
    \tilde{\mathbf{P}}\left(t\right)=e^{Jt}\mathbf{\tilde{P}}_{0}=\sum_{n=1}^{\kappa}e^{Jt}\mathbf{\tilde{P}}_{0}^{n}=\sum_{n=1}^{\kappa}e^{\bar{J}_{n}t}\mathbf{\tilde{P}}_{0}^{n}
\end{equation}
where $\bar{J}_{n}=diag\left\{ 0,0,...,0,J_{n},0...,0\right\} $ sets all blocks in $J$ to be zero except for $J_n$.
For calculating $e^{\bar{J}_{n}t}\mathbf{\tilde{P}}_{0}^{n}$ we consider the following: 
\\Let $J_0$ be a Jordan block related to an eigenvalue $\lambda_0$ with dimension $d_0$, and let $\mathbf{\Pi}$ be some vector of dimension $d_0$. Then:
\begin{equation}
    e^{J_{0}t}\mathbf{\Pi}=\sum_{i=1}^{d_{0}}\hat{e}_{i}\left(\sum_{j=1}^{d_{0}}\left(e^{J_{0}t}\right)_{ij}\Pi_{j}\right)=e^{\lambda_{0}t}\sum_{i=1}^{d_{0}}\hat{e}_{i}\left(\sum_{j=i}^{d_{0}}\frac{1}{\left(j-i\right)!}t^{j-i}\Pi_{j}\right)
\end{equation}
with $\left(\hat{e}_{i}\right)_{j}=\delta_{ij}$ being the standard basis vectors. From this we conclude: 
\begin{equation}
    \tilde{\mathbf{P}}\left(t\right)=\sum_{n=1}^{\kappa}e^{\lambda_{n}t}\sum_{i=1}^{d_{n}}\hat{e}_{i}^{n}\left(\sum_{j=i}^{d_{n}}\frac{1}{\left(j-i\right)!}t^{j-i}\left(\mathbf{\tilde{P}}_{0}^{n}\right)_{j}\right)
\end{equation}
with $\hat{e}_{i}^{n}=\hat{e}_{i+\sum_{s=1}^{n-1}d_{s}}$, a standard basis vector shifted to block $n$. 
\\In terms of $\mathbf{P}\left(t\right)
$ we use the generalized eigenvectors $\mathbf{V}_{i}^{n}=Q\hat{e}_{i}^{n}$ for block $n$ and have the complete expression:
\begin{equation}
    \mathbf{P}\left(t\right)=\sum_{n=1}^{\kappa}\sum_{i=1}^{d_{n}}e^{\lambda_{n}t}\mathbf{V}_{i}^{n}\left(\sum_{j=i}^{d_{n}}\frac{1}{\left(j-i\right)!}t^{j-i}\left(\mathbf{\tilde{P}}_{0}^{n}\right)_{j}\right)
\end{equation}
For $M$ that is given in equation (1) in the main text, there is a single eigenvalue that is $\lambda_1=0$, i.e., $d_1=1$, with corresponding eigenvector $\mathbf{V}_1^1=\mathbf{P}_{th}$. Hence: 
\begin{equation}
    \label{eqn:dynamics_sol_final_full}
    \mathbf{P}\left(t\right)=\mathbf{P}_{th}+\sum_{n=2}^{\kappa}\sum_{i=1}^{d_{n}}e^{\lambda_{n}t}\mathbf{V}_{i}^{n}\left(\sum_{j=i}^{d_{n}}\frac{1}{\left(j-i\right)!}t^{j-i}\left(\mathbf{\tilde{P}}_{0}^{n}\right)_{j}\right)
\end{equation}
In \eqref{eqn:dynamics_sol_final_full}, we sum over $N-1$ generalized eigenvectors $\mathbf{V}^n_i$ (excluding $\mathbf{P}_{th}$), where each one is multiplied by polynomial in $t$, with maximum degree of $d_n-1$. If we rename the indices by going over all generalized eigenvectors and changing $\left\{ \mathbf{V}_{i}^{n},\,\lambda_{n},\,d_{n}\right\} \to\left\{ \mathbf{V}_{s},\,\lambda_{s},\,d_{s}\right\} $ s.t. we allow repetitions in $\lambda_s$ and $d_s$ is the dimension of the block related to $\mathbf{V}_s$, then the expression can be written as:
\begin{equation}
    \mathbf{P}\left(t\right)=\mathbf{P}_{th}+\sum_{s=2}^{N}\sum_{\ell=0}^{d_{n}-1}b_{s\ell}t^{\ell}e^{\lambda_{s}t}\mathbf{V}_{s}
\end{equation}
where $b_{s\ell}$ can be zero and depend on $\mathbf{\tilde{P}}_{0}$. Therefore, $b_{s\ell}$ are related to the initial conditions.

\section{Eigen-decomposition of the transition rate matrix and condition for oscillations}
The eigenvalues of a transition rate matrix for a 3-level system are:
\begin{align}
        \begin{split}
            \lambda_{0}&=0\\\lambda_{\pm}&=-\frac{1}{2}\omega_{dis}\pm\frac{1}{2}\sqrt{\gamma}
        \end{split}
    \end{align}
    where 
    \begin{align}
    \label{eqn:eigenvalues_gamma}
        \begin{split}
            \omega_{dis}&=\sum_{i\ne j}a_{ij}\\
            \gamma&=\omega_{dis}^{2}-4\left(a_{-0}a_{-+}+a_{-+}a_{0-}+a_{-0}a_{0+}+a_{0-}a_{0+}+\right.\\&+\left.a_{-0}a_{+-}+a_{0+}a_{+-}+a_{-+}a_{+0}+a_{0-}a_{+0}+a_{+-}a_{+0}\right)
        \end{split}
    \end{align}
The corresponding eigenvectors are:
     \begin{align}
        \begin{split}
        \mathbf{V}^{0}&=\left(\begin{array}{c}
a_{-0}a_{-+}+a_{-0}a_{0+}+a_{-+}a_{+0}\\
a_{0+}a_{0-}+a_{0+}a_{+-}+a_{0-}a_{-+}\\
a_{+-}a_{+0}+a_{+-}a_{-0}+a_{+0}a_{0-}
\end{array}\right)\\
\mathbf{V}^{\pm}&=\left(\begin{array}{c}
\alpha_{1}+\alpha_{2}\pm\sqrt{\gamma}\\
\alpha_{1}-\alpha_{2}\mp\sqrt{\gamma}\\
-2\alpha_{1}
\end{array}\right)         
        \end{split}
    \end{align}
    where 
         \begin{align}
        \begin{split}
        \alpha_{1}&=a_{+0}-a_{+-}\\\alpha_{2}&=-a_{-0}+a_{-+}-a_{0-}+a_{0+}         
        \end{split}
    \end{align}
    As we can see, we have a single zero eigenvalue, $\lambda_0$. Its respective  eigenvector is proportional to the steady state vector $ \mathbf{V}^{0}\propto\mathbf{P}_{th}$.\\
    Assuming a thermal stationary state, then $\mathbf{P}_{th}$ describes the Boltzmann distribution:
    \begin{equation}
        \mathbf{P}_{th}=\frac{1}{Z_S}\left(\begin{array}{c}
e^{-\beta\mathcal{E}_{-}}\\
e^{-\beta\mathcal{E}_{0}}\\
e^{-\beta\mathcal{E}_{+}}
\end{array}\right)
    \end{equation}
With this assumption, requiring $M\mathbf{P}_{th}=0$ one gets the thermalization conditions:

\begin{align}
    \begin{split}
        a_{0-}\left(I_{-0}-1\right)+a_{+-}\left(I_{-+}-1\right)&=0\\a_{-0}\left(I_{0-}-1\right)+a_{+0}\left(I_{0+}-1\right)&=0\\a_{-+}\left(I_{+-}-1\right)+a_{0+}\left(I_{+0}-1\right)&=0
    \end{split}
\end{align}
where we define $a_{ij}e^{-\beta\mathcal{E}_{j}}=a_{ji}e^{-\beta\mathcal{E}_{i}}I_{ij}$. Noting that $I_{ij}=I^{-1}_{ji}$,  we can write:
\begin{align}
    \begin{split}
        I_{-0}&=\frac{a_{+0}\left(I_{0+}-1\right)e^{\beta\left(\mathcal{E}_{-}-\mathcal{E}_{0}\right)}}{a_{0-}}+1\\I_{+-}&=\frac{a_{+0}\left(I_{0+}-1\right)e^{\beta\left(\mathcal{E}_{+}-\mathcal{E}_{0}\right)}}{a_{-+}}+1
    \end{split}
\end{align}
Plugging these relations on $\gamma$ we get:
\begin{equation}
    \gamma=\omega_{dis}^{2}-4\frac{c}{\omega_{dis,DB}-\omega_{dis}}\sum_{i,j}e^{\beta\left(\mathcal{E}_{i}-\mathcal{E}_{j}\right)}
\end{equation}
where 
\begin{align}
    \begin{split}
        \omega_{dis}&=a_{-+}\left(e^{\beta\left(\mathcal{E}_{-}-\mathcal{E}_{+}\right)}+1\right)+a_{0-}\left(e^{\beta\left(\mathcal{E}_{0}-\mathcal{E}_{-}\right)}+1\right)+a_{+0}\left(I_{0+}\left(1+e^{\beta\left(\mathcal{E}_{-}-\mathcal{E}_{0}\right)}+e^{\beta\left(\mathcal{E}_{+}-\mathcal{E}_{0}\right)}\right)-e^{\beta\left(\mathcal{E}_{-}-\mathcal{E}_{0}\right)}\right)\\\omega_{dis,DB}&=a_{-+}\left(e^{\beta\left(\mathcal{E}_{-}-\mathcal{E}_{+}\right)}+1\right)+a_{0-}\left(e^{\beta\left(\mathcal{E}_{0}-\mathcal{E}_{-}\right)}+1\right)+a_{+0}\left(1+e^{\beta\left(\mathcal{E}_{+}-\mathcal{E}_{0}\right)}\right)
    \end{split}
\end{align}
and we use that $\text{sign}\left(c\right)=\text{sign}\left(\omega_{dis,DB}-\omega_{dis}\right)$.\\
The condition for oscillations is demanding $\gamma<0$, thus becoming equivalent to the one presented in equation (5) in the main text.
\begin{equation}
   \left|c\right|>\frac{\omega_{dis}^{2}\left|\omega_{dis,DB}-\omega_{dis}\right|}{4\sum_{i,j}e^{\beta\left(\mathcal{E}_{i}-\mathcal{E}_{j}\right)}}=\frac{\omega_{dis}^{2}\left|\omega_{dis,DB}-\omega_{dis}\right|}{4\left(3+\sum_{i\ne j}e^{\beta\left(\mathcal{E}_{i}-\mathcal{E}_{j}\right)}\right)} 
\end{equation}
Note that for oscillations with $\gamma<0$, the oscillations number during the system thermalization time scale is defined as the ratio between the imaginary and real part of the non-zero eigenvalues, i.e., $|\sqrt{\gamma}/\omega_{dis}|$. We can bound the square of this quantity by taking only the negative terms in $\gamma$:
\begin{equation}
    \frac{\left|\gamma\right|}{\omega_{dis}^{2}}<2\frac{a_{-0}\left(a_{-+}+a_{0+}+a_{+-}\right)+a_{-+}\left(a_{0-}+a_{+0}\right)+\left(a_{0-}+a_{+-}\right)\left(a_{0+}+a_{+0}\right)}{\left(\sum_{i\ne j}a_{ij}\right)^{2}}<1
\end{equation}

\section{Derivation of  rates of transitions and the T matrix elements}

\subsection{The Low-Density Limit in Open Quantum Systems}
In the low-density limit the quantum master equation is derived for a discrete level system coupled to a thermal bath of free particles. The local interaction between the gas and the system describes scattering processes where the gas density is taken to be low enough such that we can consider each scattering process to be independent from one another. Therefore, these processes involve only a single particle scattering process.\\
The effective Hamiltonian is:
\begin{equation}
\label{eqn:eff_Ham_full}
    H=H_{P}\otimes 1 +1\otimes H_{S}+H_{int}
\end{equation}
where
\begin{align}
\label{eqn:part_syst_Ham}
    \begin{split}
        H_{P}&=\intop d\boldsymbol{p}E_{\boldsymbol{p}}\left|\boldsymbol{p}\right\rangle \left\langle \boldsymbol{p}\right|\\H_{S}&=\sum_{j}\mathcal{E}_{j}\left|j\right\rangle \left\langle j\right|
    \end{split}
\end{align}
$H_P$ being the free particle Hamiltonian, with $E_{\boldsymbol{p}}=\frac{\boldsymbol{p}^2}{2m}$, and $H_S$ being the system Hamiltonian. We denote also $H_0=H_{P}\otimes 1 +1\otimes H_{S}$ for later use.\\
When deriving the quantum master equation, the particle's momentum distribution is given by Boltzmann distribution:
\begin{equation}
\label{eqn:mom_dist}
    G\left(\boldsymbol{p}\right)=\frac{e^{-\beta E_{\boldsymbol{p}}}}{Z_{P}}
\end{equation}
The master equation then takes the form:
\begin{equation}
\label{eqn:master_eqn}
    \frac{d}{dt}\rho_{S}=\left(\mathcal{L}_{S}+\mathcal{L}_{D}\right)\rho_{S}      
\end{equation}
where $\mathcal{L}_{S}=-i\left[H_{S},\cdot\right]$ is the system's Liouvillian, describing the system's unitary evolution, and $\mathcal{L}_{D}$ is the dissipator:
\begin{align}
\label{eqn:dissipator}
    \begin{split}           \mathcal{L}_{D}\rho_{S}&=\nu\pi\sum_{\omega\in\text{Sp}\left(i\mathcal{L}_{S}\right)}\intop d\boldsymbol{p}d\boldsymbol{p}'G\left(\boldsymbol{p}\right)\delta\left(E_{\boldsymbol{p}'}-E_{\boldsymbol{p}}+\omega\right)\left(\left[T_{\omega}\left(\boldsymbol{p}',\boldsymbol{p}\right)\rho_{S},T_{\omega}^{\dagger}\left(\boldsymbol{p}',\boldsymbol{p}\right)\right]\right.\\&+\left.\left[T_{\omega}\left(\boldsymbol{p}',\boldsymbol{p}\right),\rho_{S}T_{\omega}^{\dagger}\left(\boldsymbol{p}',\boldsymbol{p}\right)\right]\right)
    \end{split}
\end{align}
    where $\text{Sp}\left(i\mathcal{L}_{S}\right)$ is the spectrum of $i\mathcal{L}_{S}$ (all possible eigenenergies differences).
    The jump operators $T_{\omega}\left(\boldsymbol{p}',\boldsymbol{p}\right)$ are given by:
    \begin{equation}        
    \label{eqn:T_opertors}T_{\omega}\left(\boldsymbol{p}',\boldsymbol{p}\right)=\sum_{\mathcal{E}_{k}-\mathcal{E}_{\ell}=\omega}\left\langle \boldsymbol{p}',k\left|T\right|\boldsymbol{p},\ell\right\rangle \left|k\right\rangle \left\langle \ell\right|
    \end{equation}
    with $\left\langle \boldsymbol{p}',k\left|T\right|\boldsymbol{p},\ell\right\rangle $ being the T-matrix element related to the scattering process $\left|\boldsymbol{p},\ell\right\rangle \to\left|\boldsymbol{p}',k\right\rangle$. For a non-degenerate $H_S$, the populations $P_{i}=\left\langle i\left|\rho_{S}\right|i\right\rangle$ are governed by the Pauli rate equation with transition rates given by equation (4) in the main text.
\subsection{T matrix elements}
The T-matrix can be defined as:
    \begin{equation}
    \label{eqn:T_def}T\left(E\right)=H_{int}+H_{int}G\left(E\right)H_{int}
    \end{equation}
    with $G\left(E\right)=\left(E-H\right)^{-1}$ being the Green operator of the entire Hamiltonian.
    Equivalently, we can write the Lippmann-Schwinger equation for $T$:
    \begin{equation} 
    \label{eqn:T_LS}
    T\left(E\right)=H_{int}+H_{int}G_{0}\left(E\right)T\left(E\right)
    \end{equation}
    with $G_0\left(E\right)=\left(E-H_0\right)^{-1}$ being the Green operator of the free Hamiltonian. By introducing the Moller operators:
    \begin{equation}
    \label{eqn:Moller_def}\Omega_{\pm}=\lim_{t\to\mp\infty}U^{\dagger}\left(t\right)U_{0}\left(t\right)=\lim_{t\to\mp\infty}e^{iHt}e^{-iH_{0}t}
    \end{equation}
    we map the asymptotic states to the state of the system at time 0:
    \begin{align}
        \begin{split}
            \label{eqn:Moller_states}\Omega_{+}\left|\psi_{in}\right\rangle &=\Omega_{+}\left|\boldsymbol{p},\ell\right\rangle =\left|\psi\right\rangle \\\Omega_{-}\left|\psi_{out}\right\rangle &=\Omega_{-}\left|\boldsymbol{p}',k\right\rangle =\left|\psi\right\rangle         
        \end{split}
    \end{align}
This allows us to write the T matrix elements as: 
\begin{equation}
     \left\langle \boldsymbol{p}',k\left|T\right|\boldsymbol{p},\ell\right\rangle =\left\langle \boldsymbol{p}',k\left|H_{int}\right|\boldsymbol{p},\ell\right\rangle +\left\langle \boldsymbol{p}',k\left|H_{int}G_{0}H_{int}\right|\psi\right\rangle 
\end{equation}
with the energy being $E=E_{\textbf{p}}+\mathcal{E}_\ell$.\\
    The state $\left|\psi\right\rangle$ can be written as
    \begin{equation}
        \left|\psi\right\rangle =\Omega_{+}\left|\boldsymbol{p},\ell\right\rangle =\left|\boldsymbol{p},\ell\right\rangle +G_{0}H_{int}\left|\psi\right\rangle 
    \end{equation}
    such that 
    \begin{equation}
        \label{eqn:T-LS-explicit_final}
        \left\langle \boldsymbol{p}',k\left|T\right|\boldsymbol{p},\ell\right\rangle =\left\langle \boldsymbol{p}',k\left|H_{int}\right|\psi\right\rangle 
    \end{equation}
\subsection{Toy model T matrix elements}
The interaction term is given by:
\begin{equation}
\label{eqn:inter_general}        H_{int}=\sum_{i}V_{i}\left(\hat{\boldsymbol{q}}\right)\left|\chi_{i}\right\rangle \left\langle \chi_{i}\right|
    \end{equation}
    where $V_i(\textbf{q})$ are the scattering potential in each site $i$, and  $\left|\chi_{i}\right\rangle $ is in the position basis of the system.\\
    The T matrix elements take the form:
\begin{align}
    \begin{split}
    \label{eqn:T_elemnt_state_calc}
        \left\langle \boldsymbol{p}'j'\left|T\right|\boldsymbol{p}j\right\rangle =&\left\langle \boldsymbol{p}'j'\left|H_{int}\right|\psi\right\rangle \\=&\left\langle \boldsymbol{p}'j'\left|\left(\sum_{i}V_{i}\left(\hat{\boldsymbol{q}}\right)\left|\chi_{i}\right\rangle \left\langle \chi_{i}\right|\right)\intop_{\mathbb{R}^{d}}\mathrm{~d}^{d}q'\sum_{k}\psi_{k}\left(\boldsymbol{q}'\right)\right|\boldsymbol{q}'k\right\rangle =\\=&\sum_{k}\sum_{i}\left(\left\langle j'\mid\chi_{i}\right\rangle \left\langle \chi_{i}\mid k\right\rangle \right)\intop_{\mathbb{R}^{d}}\mathrm{~d}^{d}q'\psi_{k}\left(\boldsymbol{q}'\right)\left\langle \boldsymbol{p}'\left|V_{i}\left(\hat{\boldsymbol{q}}\right)\right|\boldsymbol{q}'\right\rangle =\\=&\sum_{k}\sum_{i}\left(\left\langle j'\mid\chi_{i}\right\rangle \left\langle \chi_{i}\mid k\right\rangle \right)\intop_{\mathbb{R}^{d}}\mathrm{~d}^{d}q'\psi_{k}\left(\boldsymbol{q}'\right)\frac{e^{-\frac{i}{\hbar}\boldsymbol{p}'\cdot\boldsymbol{q'}}}{\left(2\pi\hbar\right)^{d/2}}V_{i}\left(\boldsymbol{q}'\right)
    \end{split} 
\end{align}
where we used:
\begin{equation}
    \psi_{j}(\boldsymbol{q})=\langle\boldsymbol{q}j\mid\psi\rangle,\quad\tilde{\psi}_{j}(\boldsymbol{p})=\langle\boldsymbol{p}j\mid\psi\rangle
\end{equation}
and:
\begin{equation}
    \left\langle \boldsymbol{q}j\mid\boldsymbol{p}j'\right\rangle =\delta_{jj'}\frac{e^{\frac{i}{\hbar}\boldsymbol{p}\cdot\boldsymbol{q}}}{\left(2\pi\hbar\right)^{d/2}}
\end{equation}
The wavefunction in momentum space is given by:
\begin{align}
    \begin{split}
        \label{eqn:state_momentum}\tilde{\psi}_{j'}\left(\boldsymbol{p}'\right)&=\left\langle \boldsymbol{p}'j'\mid\boldsymbol{p}j\right\rangle +\left\langle \boldsymbol{p}'j'\left|G_{0}H_{int}\right|\psi\right\rangle =\\&=\delta^{d}\left(\boldsymbol{p}-\boldsymbol{p}'\right)\delta_{jj'}+\frac{\left\langle \boldsymbol{p}'j'\left|H_{int}\right|\psi\right\rangle }{E-E_{\boldsymbol{p}'}-\mathcal{E}_{j'}+i\varepsilon}
    \end{split}
\end{align}
where we use the geometric series expansion:
\begin{equation}
    \left\langle \boldsymbol{p}'j'\right|G_{0}\left(E+i\varepsilon\right)=\left\langle \boldsymbol{p}'j'\right|\frac{1}{E-H_{0}+i\varepsilon}=\left\langle \boldsymbol{p}'j'\right|\frac{1}{E-E_{\boldsymbol{p}'}-\mathcal{E}_{j'}+i\varepsilon}
\end{equation}
This allows us to write a closed equation for the wavefunction:
\begin{align}
    \begin{split}
        \label{eqn:state_momentum_final}
        \tilde{\psi}_{j'}\left(\boldsymbol{p}'\right)&=\delta^{d}\left(\boldsymbol{p}-\boldsymbol{p}'\right)\delta_{jj'}+\\&+\frac{1}{E-E_{\boldsymbol{p}'}-\mathcal{E}_{j'}+i\varepsilon}\sum_{k}\sum_{i}\left(\left\langle j'\mid\chi_{i}\right\rangle \left\langle \chi_{i}\mid k\right\rangle \right)\intop_{\mathbb{R}^{d}}\mathrm{~d}^{d}q'\psi_{k}\left(\boldsymbol{q}'\right)\frac{e^{-\frac{i}{\hbar}\boldsymbol{p}'\cdot\boldsymbol{q'}}}{\left(2\pi\hbar\right)^{d}}V_{i}\left(\boldsymbol{q}'\right)
    \end{split}
\end{align}
and by performing the Fourier transform on both sides:
\small
\begin{align}
    \begin{split}
        \label{eqn:state_position_final}
        \psi_{j'}\left(\boldsymbol{q}\right)=&\intop_{\mathbb{R}^{d}}\mathrm{~d}^{d}p'\frac{e^{\frac{i}{\hbar}\boldsymbol{p}'\cdot\boldsymbol{q}}}{\left(2\pi\hbar\right)^{d/2}}\tilde{\psi}_{j'}\left(\boldsymbol{p}'\right)=\intop_{\mathbb{R}^{d}}\mathrm{~d}^{d}p'\frac{e^{\frac{i}{\hbar}\boldsymbol{p}'\cdot\boldsymbol{q}}}{\left(2\pi\hbar\right)^{d/2}}\delta^{d}\left(\boldsymbol{p}-\boldsymbol{p}'\right)\delta_{jj'}+\\+&\intop_{\mathbb{R}^{d}}\mathrm{~d}^{d}p'\frac{e^{\frac{i}{\hbar}\boldsymbol{p}'\cdot\boldsymbol{q}}}{\left(2\pi\hbar\right)^{d/2}}\frac{1}{E-E_{\boldsymbol{p}'}-\mathcal{E}_{j'}+i\varepsilon}\sum_{k}\sum_{i}\left(\left\langle j'\mid\chi_{i}\right\rangle \left\langle \chi_{i}\mid k\right\rangle \right)\intop_{\mathbb{R}^{d}}\mathrm{~d}^{d}q'\psi_{k}\left(\boldsymbol{q}'\right)\frac{e^{-\frac{i}{\hbar}\boldsymbol{p}'\cdot\boldsymbol{q'}}}{\left(2\pi\hbar\right)^{d/2}}V_{i}\left(\boldsymbol{q}'\right)=\\=&\frac{e^{\frac{i}{\hbar}\boldsymbol{p}\cdot\boldsymbol{q}}}{\left(2\pi\hbar\right)^{d/2}}\delta_{jj'}+\\+&\sum_{k}\sum_{i}\left(\left\langle j'\mid\chi_{i}\right\rangle \left\langle \chi_{i}\mid k\right\rangle \right)\intop_{\mathbb{R}^{d}}\mathrm{~d}^{d}p'\frac{1}{E-E_{\boldsymbol{p}'}-\mathcal{E}_{j'}+i\varepsilon}\intop_{\mathbb{R}^{d}}\mathrm{~d}^{d}q'\psi_{k}\left(\boldsymbol{q}'\right)\frac{e^{-\frac{i}{\hbar}\boldsymbol{p}'\cdot\left(\boldsymbol{q'-q}\right)}}{\left(2\pi\hbar\right)^{d}}V_{i}\left(\boldsymbol{q}'\right)
    \end{split}
\end{align}
\normalsize
Taking a few simplifications for the model:
\begin{enumerate}
    \item \textbf{Delta interaction}: Introduced as $H_{int}=\sum_{i\in \{1,2,3\}}V_{i}\delta(\mathbf{q}-\mathbf{q_{i}})|\chi_i\rangle\langle \chi_i|$. Choosing $\textbf{q}=\textbf{q}_{i'}$, i.e. in the position of the sites, gives:
\small
\begin{align}
\label{eqn:state_position_delta} 
    \begin{split}         \psi_{j'}\left(\boldsymbol{q}_{i'}\right)=&\frac{e^{\frac{i}{\hbar}\boldsymbol{p}\cdot\boldsymbol{q}_{i'}}}{\left(2\pi\hbar\right)^{d/2}}\delta_{jj'}+\\+&\sum_{k}\sum_{i}\left(\mathcal{V}_{i}\left\langle j'\mid\chi_{i}\right\rangle \left\langle \chi_{i}\mid k\right\rangle \right)\psi_{k}\left(\boldsymbol{q}_{i}\right)\intop_{\mathbb{R}^{d}}\mathrm{~d}^{d}p'\frac{1}{E_{\boldsymbol{p}}+\mathcal{E}_{j}-E_{\boldsymbol{p}'}-\mathcal{E}_{j'}+i\varepsilon}\frac{e^{-\frac{i}{\hbar}\boldsymbol{p}'\cdot\left(\boldsymbol{q}_{i}-\boldsymbol{q}_{i'}\right)}}{\left(2\pi\hbar\right)^{d}}
    \end{split}
\end{align}
\normalsize    

where we used that $E=E_{\textbf{p}}+\mathcal{E}_j$.
The equation for the wavefunction has now become a set of $N^2$ linear equations, that for each $j$ and given some $\textbf{p}$, we solve for the vector $\vec{\psi}\left(\boldsymbol{q}_{i'}\right)$. Having the solutions at hand, we plug them into the T matrix elements equation:
\begin{align}
    \begin{split}
        \left\langle \boldsymbol{p}'j'\left|T\right|\boldsymbol{p}j\right\rangle &=\sum_{k}\sum_{i}\left(\left\langle j'\mid\chi_{i}\right\rangle \left\langle \chi_{i}\mid k\right\rangle \right)\intop_{\mathbb{R}^{d}}\mathrm{~d}^{d}q'\psi_{k}\left(\boldsymbol{q}'\right)\frac{e^{-\frac{i}{\hbar}\boldsymbol{p}'\cdot\boldsymbol{q'}}}{\left(2\pi\hbar\right)^{d/2}}\mathcal{V}_{i}\delta\left(\boldsymbol{q}'-\boldsymbol{q}_{i}\right)=\\&=\sum_{k}\sum_{i}\left(\mathcal{V}_{i}\left\langle j'\mid\chi_{i}\right\rangle \left\langle \chi_{i}\mid k\right\rangle \right)\psi_{k}\left(\boldsymbol{q}_{i}\right)\frac{e^{-\frac{i}{\hbar}\boldsymbol{p}'\cdot\boldsymbol{q}_{i}}}{\left(2\pi\hbar\right)^{d/2}}
    \end{split}
\end{align}
\item \textbf{Short separation}: Choosing $\textbf{q}_i=0$ gives:
\begin{align}
    \begin{split}
        \psi_{j'}&=\frac{1}{\left(2\pi\hbar\right)^{d/2}}\delta_{jj'}+\frac{1}{\left(2\pi\hbar\right)^{d}}\sum_{k}v_{j'k}\psi_{k}\intop_{\mathbb{R}^{d}}\mathrm{~d}^{d}p'\frac{1}{E_{\boldsymbol{p}}+\mathcal{E}_{j}-E_{\boldsymbol{p}'}-\mathcal{E}_{j'}+i\varepsilon}=\\&=\frac{1}{\left(2\pi\hbar\right)^{d/2}}\delta_{jj'}+\frac{1}{\left(2\pi\hbar\right)^{d}}\left(v\vec{\psi}\right)_{j'}\intop_{\mathbb{R}^{d}}\mathrm{~d}^{d}p'\frac{1}{E_{\boldsymbol{p}}+\mathcal{E}_{j}-E_{\boldsymbol{p}'}-\mathcal{E}_{j'}+i\varepsilon}
    \end{split}
\end{align}
where we write $\psi_{j'}=\psi_{j'}\left(0\right)$ and define the interaction matrix $v$ and as in the main text $v_{kl}=\sum_{i=1}^{3}\langle k|\chi_i\rangle V_{i}\langle \chi_i|l\rangle$,
$k,l\in\{+,0,-\}$. This reduces the number of equations by a factor of $N$ - the number of sites. 
\item \textbf{One dimension}: Taking $d=1$ we can evaluate the integral using Cauchy's integral formula (without the renormalization required for higher dimensions), closing a contour in the upper half of the complex plane, and then taking $\varepsilon\to 0^+$:
\begin{equation}
    \intop_{-\infty}^{\infty}dp'
    \frac{1}{E-E_{p'}-\mathcal{E}_{j'}+i\varepsilon}=\intop_{-\infty}^{\infty}dp'\frac{2m}{2m\left(E-\mathcal{E}_{j'}\right)-p^{2}+i\varepsilon}=-\frac{i\pi\sqrt{2m}}{\sqrt{E-\mathcal{E}_{j'}}}
\end{equation}
\end{enumerate}
Finally, the T-matrix elements take the form:
\begin{equation}
\label{eqn:T_mat_elem1}
    \left\langle p'j'\left|T\left(E\right)\right|pj\right\rangle =\frac{1}{\sqrt{2\pi\hbar}}\sum_{k}v_{j'k}\psi_{k}\left(E\right)
\end{equation}
where $\psi$ is given by:
\begin{equation}
\label{eqn:psi_indx}
    \psi_{j'}\left(E\right)=\frac{1}{\sqrt{2\pi\hbar}}\delta_{jj'}-\frac{1}{2\pi\hbar}\frac{i\pi\sqrt{2m}}{\sqrt{E-\mathcal{E}_{j'}}}\sum_{k}v_{j'k}\psi_{k}\left(E\right)
\end{equation}
thus having:
\begin{equation}
\label{eqn:T_mat_elem2}
    \left\langle p'j'\left|T\left(E\right)\right|pj\right\rangle =i\frac{\sqrt{E-\mathcal{E}_{j'}}}{\pi\sqrt{2m}}\left(\sqrt{2\pi\hbar}\psi_{j'}\left(E\right)-\delta_{jj'}\right)
\end{equation}
The interaction matrix $v$ for the toy model is given by:
   \begin{equation}
        v_{k\ell}=\sum_{i}\mathcal{V}_{i}\left\langle k\mid\chi_{i}\right\rangle \left\langle \chi_{i}\mid\ell\right\rangle =\begin{cases}
w & k=\ell\\
u & \left(k,\ell\right)=\left(-,0\right),\left(0,+\right),\left(+,-\right)\\
u^{*} & \left(k,\ell\right)=\left(-,+\right),\left(+,0\right),\left(0,-\right)
\end{cases}
    \end{equation}
    with
    \begin{align}
        \begin{split}
            w&=\frac{1}{3}\left(\mathcal{V}_{1}+\mathcal{V}_{2}+\mathcal{V}_{3}\right)\\u&=\frac{1}{3}\left(\mathcal{V}_{1}+\mathcal{V}_{2}e^{i\frac{2\pi}{3}}+\mathcal{V}_{3}e^{-i\frac{2\pi}{3}}\right)
        \end{split}
    \end{align}

The equation for the wave function can be written in matrix form. This is allowed since the vector $\psi_{j'}$  is computed separately for any $j$. Therefore, we can define the matrix $\Psi$ with both indices that satisfies the following:
\begin{align}
    \label{eqn:psi_mat}
    \Psi_{j'j}\left(E\right)&=\frac{1}{\sqrt{2\pi\hbar}}\delta_{j'j}-\frac{i}{2\hbar}\frac{\sqrt{2m}}{\sqrt{E-\mathcal{E}_{j'}}}\sum_{k}v_{j'k}\Psi_{kj}\left(E\right)=\frac{\delta_{j'j}}{\sqrt{2\pi\hbar}}+\left(D_{1}\left(E\right)v\Psi\left(E\right)\right)_{j'j}\\\Rightarrow\Psi\left(E\right)&=\frac{1}{\sqrt{2\pi\hbar}}\left(1-D_{1}\left(E\right)v\right)^{-1}
\end{align}
where we define $\left(D_{1}\left(E\right)\right)_{j'j}=-\frac{i}{2\hbar}\frac{\sqrt{2m}}{\sqrt{E-\mathcal{E}_{j'}}}\delta_{j'j}$.\\
The T matrix is given by:
\begin{equation}        T\left(E\right)=D_{2}\left(E\right)\left(\left(1-D_{1}\left(E\right)v\right)^{-1}-1\right)
\end{equation}
where $T_{j'j}\left(E\right)=\left\langle p'j'\left|T\left(E\right)\right|pj\right\rangle $ and  $\left(D_{2}\left(E\right)\right)_{j'j}=\frac{i}{\pi}\frac{\sqrt{E-\mathcal{E}_{j'}}}{\sqrt{2m}}\delta_{j'j}$.

\subsection{Explicit T matrix elements}
The explicit form is $T_{ij}\left(E\right)=\frac{\tilde{T}_{ij}\left(E\right)}{D_T\left(E\right)}$ where:
\begin{align}
    \begin{split}
        D_{T}\left(E\right)&=i\pi m\sqrt{2m}\left(2\left(3\left(\Im\left(u\right)\right)^{2}\Re\left(u\right)-\left(\Re\left(u\right)\right)^{3}\right)+3\left|u\right|^{2}w-w^{3}\right)+\\&+2\pi m\hbar\left(\left|u\right|^{2}-w^{2}\right)\left(\sqrt{E-\mathcal{E}_{-}}+\sqrt{E-\mathcal{E}_{0}}+\sqrt{E-\mathcal{E}_{+}}\right)+\\&+2i\pi\hbar^{2}\sqrt{2m}w\left(\sqrt{E-\mathcal{E}_{-}}\sqrt{E-\mathcal{E}_{0}}+\sqrt{E-\mathcal{E}_{-}}\sqrt{E-\mathcal{E}_{+}}+\sqrt{E-\mathcal{E}_{0}}\sqrt{E-\mathcal{E}_{+}}\right)+\\&+4\pi\hbar^{3}\sqrt{E-\mathcal{E}_{-}}\sqrt{E-\mathcal{E}_{0}}\sqrt{E-\mathcal{E}_{+}}
    \end{split}
\end{align}
and:
\begin{equation}
\tilde{T}_{ij}\left(E\right)=\begin{cases}
\sqrt{2}\hbar\sqrt{E-\mathcal{E}_{i}}\sqrt{E-\mathcal{E}_{j}}\left(u^{*}\sqrt{2}\hbar\sqrt{E-\mathcal{E}_{k\ne i,j}}+i\sqrt{m}\left(u^{*}w-u^{2}\right)\right) & q=0\\
\sqrt{2}\hbar\sqrt{E-\mathcal{E}_{i}}\sqrt{E-\mathcal{E}_{j}}\left(u\sqrt{2}\hbar\sqrt{E-\mathcal{E}_{k\ne i,j}}+i\sqrt{m}\left(uw-\left(u^{*}\right)^{2}\right)\right) & q=1
\end{cases}
\end{equation}
with $q=0$ for clockwise rates and $q=1$ for counterclockwise rates.
Since we are interested in $\left|T_{ij}\left(E\right)\right|^{2}$, we look at $\left|\tilde{T}_{ij}\left(E\right)\right|^{2}$, but we should consider the value of which $E$ takes, since $E<\mathcal{E}_{+}$ implies $\sqrt{E-\mathcal{E}_{+}}\in i\mathbb{R}$:
\\\underline{$E>\mathcal{E}_{+}$}:
\begin{align}
    \begin{split}        \left|\tilde{T}_{ij}\left(E\right)\right|^{2}&=2\hbar^{2}\left(E-\mathcal{E}_{i}\right)\left(E-\mathcal{E}_{j}\right)\left(2\left|u\right|^{2}\hbar^{2}\left(E-\mathcal{E}_{k\ne i,j}\right)+m\left|u^{*}w-u^{2}\right|^{2}\right.\\&\left.+\left(-1\right)^{q_{ij}}2\hbar\sqrt{2m}\sqrt{E-\mathcal{E}_{k\ne i,j}}\Im\left(u^{3}\right)\right)
    \end{split}
\end{align}
\underline{$\mathcal{E}_{0}<E<\mathcal{E}_{+}$}: (we will see later this is relevant only for rates $a_{0-},a_{-0}$)
\begin{align}
    \begin{split}       \left|\tilde{T}_{-0}\left(E\right)\right|^{2}=\left|\tilde{T}_{0-}\left(E\right)\right|^{2}&=2\hbar^{2}\left(E-\mathcal{E}_{-}\right)\left(E-\mathcal{E}_{0}\right)\left(2\left|u\right|^{2}\hbar^{2}\left(\mathcal{E}_{+}-E\right)+m\left|u^{*}w-u^{2}\right|^{2}\right.\\&\left.+2\hbar\sqrt{2m}\sqrt{\mathcal{E}_{+}-E}\left(2\left|u\right|^{2}w-\Re\left(u^{3}\right)\right)\right)
    \end{split}
\end{align}
\subsection{Violation of micro-reversibility}
Micro-reversibility is governed by the difference (non zero for $E>\mathcal{E}_+$):
\begin{gather}
    \left|\left\langle \boldsymbol{p}',i\left|T\left(E\right)\right|\boldsymbol{p},j\right\rangle \right|^{2}-\left|\left\langle -\boldsymbol{p},j\left|T\left(E\right)\right|-\boldsymbol{p}',i\right\rangle \right|^{2}=\nonumber\\=\left|\left\langle \boldsymbol{p}',i\left|T\left(E\right)\right|\boldsymbol{p},j\right\rangle \right|^{2}-\left|\left\langle \boldsymbol{p},j\left|T\left(E\right)\right|\boldsymbol{p}',i\right\rangle \right|^{2}=\nonumber\\=\left(-1\right)^{q}\frac{\left(2\hbar\right)^{3}\sqrt{2m}\left(E-\mathcal{E}_{i}\right)\left(E-\mathcal{E}_{j}\right)\sqrt{E-\mathcal{E}_{k\ne i,j}}}{\left|D_{T}\left(E\right)\right|^{2}}\Im\left(u^{3}\right)
\end{gather}
where $6\sqrt{3}\Im\left(u^{3}\right)=\left(\mathcal{V}_{1}-\mathcal{V}_{2}\right)\left(\mathcal{V}_{1}-\mathcal{V}_{3}\right)\left(\mathcal{V}_{2}-\mathcal{V}_{3}\right)$.

\section{Decomposition of the rates of transition}
Since the T matrix elements are computed for the on-shell energy of the ingoing state $E=E_p+\mathcal{E}_\ell$ (with $E_p=\frac{p^2}{2m}$), we can change the integration variable from $p$ to $E$:
\begin{align}
    \begin{split}
        a_{k\ell}=&2\nu\pi\intop dpdp'\frac{e^{-\beta E_{p}}}{Z_P}\delta\left(E_{p'}+\mathcal{E}_{k}-E_{p}-\mathcal{E}_{\ell}\right)\left|\left\langle k,p'\left|T\right|\ell,p\right\rangle \right|^{2}\\=&\frac{2\pi\nu}{Z_P}\intop_{\mathcal{E}_{\ell}}^{\infty}dE\intop_{-\infty}^{\infty}dp'\,e^{-\beta\left(E-\mathcal{E}_{\ell}\right)}\frac{m}{\sqrt{2m\left(E-\mathcal{E}_{\ell}\right)}}\delta\left(E_{p'}+\mathcal{E}_{k}-E\right)\left|\left\langle p',k\left|T\right|p,\ell\right\rangle \right|^{2}=\\=&\frac{2\pi\nu}{Z_P}\intop_{\mathcal{E}_{\ell}}^{\infty}dE\intop_{-\infty}^{\infty}dp'e^{-\beta\left(E-\mathcal{E}_{\ell}\right)}\frac{m}{\sqrt{2m\left(E-\mathcal{E}_{\ell}\right)}}2m\delta\left(p'^{2}-2m\left(E-\mathcal{E}_{k}\right)\right)\left|\left\langle p',k\left|T\right|p,\ell\right\rangle \right|^{2}=\\=&\frac{2\pi\nu}{Z_{P}}\intop_{\mathcal{E}_{\ell}}^{\infty}dE\intop_{-\infty}^{\infty}dp'e^{-\beta\left(E-\mathcal{E}_{\ell}\right)}\frac{2m^{2}}{\sqrt{2m\left(E-\mathcal{E}_{\ell}\right)}}\sum_{\alpha=\pm1}\left(\frac{\delta\left(p'+\alpha\sqrt{2m\left(E-\mathcal{E}_{k}\right)}\right)}{\sqrt{2m\left(E-\mathcal{E}_{k}\right)}}\right)\left|\left\langle p',k\left|T\right|p,\ell\right\rangle \right|^{2}
    \end{split}
\end{align}
The term inside the $\delta$ function implies $E>\mathcal{E}_k$, thus integration over $p'$ gives:
\begin{equation}
\label{eqn:rates_E}
    a_{k\ell}=\frac{4\pi m\nu}{Z_P}\intop_{\max\{\mathcal{E}_{k},\mathcal{E}_{\ell}\}}^{\infty}dEe^{-\beta\left(E-\mathcal{E}_{\ell}\right)}\frac{\left|T_{k\ell}\left(E\right)\right|^{2}}{\sqrt{E-\mathcal{E}_{k}}\sqrt{E-\mathcal{E}_{\ell}}}
\end{equation}
We recall that inside the integration, $p,p'$ are defined by $E$ and $\ell,k$, respectively. Hence we can write the T matrix as $T_{k\ell}\left(E\right)\equiv\left\langle p',k\left|T\left(E\right)\right|p,\ell\right\rangle$. Note that the partition function for a single free particle has: $Z_P^{-1}\propto\sqrt{\beta}$.
 \\Denoting $\tilde{Z}_P=Z_P\sqrt{\beta}$, we have that $\tilde{Z}_P$ is independent of temprature. Thus, we define:
 \begin{align}
     \begin{split}
        a_{0}=&\frac{4\pi m\nu}{\tilde{Z}_{P}}\frac{2\hbar^{2}\sqrt{\left(E-\mathcal{E}_{+}\right)\left(E-\mathcal{E}_{0}\right)\left(E-\mathcal{E}_{-}\right)}}{\left|D_{T}\left(E\right)\right|^{2}}\left(2\left|u\right|^{2}\hbar^{2}\sqrt{E}+m\left|u^{*}w-u^{2}\right|^{2}\frac{1}{\sqrt{E}}\right)\\&\\a_{1}=&\frac{4\pi m\nu}{\tilde{Z}_{P}}\frac{2\hbar^{2}\sqrt{\left(E-\mathcal{E}_{+}\right)\left(E-\mathcal{E}_{0}\right)\left(E-\mathcal{E}_{-}\right)}}{\left|D_{T}\left(E\right)\right|^{2}}\left(2\sqrt{2m}\hbar\Im\left(u^{3}\right)\right)\\&\\b_{k\ell}=&\frac{4\pi m\nu}{\tilde{Z}_{P}}\frac{2\hbar^{2}\sqrt{\left(E-\mathcal{E}_{+}\right)\left(E-\mathcal{E}_{0}\right)\left(E-\mathcal{E}_{-}\right)}}{\left|D_{T}\left(E\right)\right|^{2}}\times\\&\left(2\left|u\right|^{2}\hbar^{2}\left(\sqrt{E-\mathcal{E}_{n\ne k,\ell}}-\sqrt{E}\right)+m\left|u^{*}w-u^{2}\right|^{2}\left(\frac{1}{\sqrt{E-\mathcal{E}_{n\ne k,\ell}}}-\frac{1}{\sqrt{E}}\right)\right)\\\tilde{b}_{k\ell}=&\begin{cases}
\frac{4\pi m\nu}{\tilde{Z}_{P}}\frac{1}{\sqrt{\left(E-\mathcal{E}_{-}\right)\left(E-\mathcal{E}_{0}\right)}}\frac{\left|\tilde{T}_{12}\left(E\right)\right|^{2}}{\left|D_{T}\left(E\right)\right|^{2}} & k,\ell\in\left\{ 0,-\right\} \\
0 & else
\end{cases}
     \end{split}
 \end{align}
with these definitions, one arrives at the main text's expression in equation (6). Note that at high energies:
\begin{equation}
    \frac{\sqrt{\left(E-\mathcal{E}_{+}\right)\left(E-\mathcal{E}_{0}\right)\left(E-\mathcal{E}_{-}\right)}}{\left|D_{T}\left(E\right)\right|^{2}}\approx\frac{1}{\left(4\pi\hbar^{3}\right)^{2}E^{\frac{3}{2}}}
\end{equation}
So, in leading order:
\begin{align}
    \begin{split}
        a_{0}&\propto\frac{\left|u\right|^{2}}{E}\\a_{1}&\propto\frac{\Im\left(u^{3}\right)}{E^{\frac{3}{2}}}\\b_{k\ell}&\propto\frac{\left|u\right|^{2}\mathcal{E}_{n\ne k,\ell}}{E^{2}}
    \end{split}
\end{align}
Implying equation (7) in the main paper.

Additionally, with this decomposition we can write:
\begin{equation}
    \tilde{a}_{k\ell}=\sqrt{\beta}\left(\int_{\mathcal{E}_{+}}^{\infty}dEe^{-\beta E}\left(a_{0}+b_{k\ell}\right)+\int_{\mathcal{E}_{0}}^{\mathcal{E}_{+}}dEe^{-\beta E}\tilde{b}_{k\ell}\right)
\end{equation}
such that $a_{k\ell}=\tilde{a}_{k\ell}+\left(-1\right)^{q}\tilde{a}_{1}$ where $\tilde{a}_{1}=\sqrt{\beta}\int_{\mathcal{E}_{+}}^{\infty}dEe^{-\beta E}a_{1}\propto\Delta_{VMR}$,  and $\tilde{a}_{k\ell}=\tilde{a}_{\ell k}>0$ by definition. Hence, we can write:
\begin{align}
    \begin{split}
    c&=a_{-+}a_{+0}a_{0-}-a_{-0}a_{0+}a_{+-}=\\&=e^{\beta\sum_{\ell}\mathcal{E}_{\ell}}\left(\left(\tilde{a}_{-+}+\tilde{a}_{1}\right)\left(\tilde{a}_{+0}+\tilde{a}_{1}\right)\left(\tilde{a}_{0-}+\tilde{a}_{1}\right)-\left(\tilde{a}_{-0}-\tilde{a}_{1}\right)\left(\tilde{a}_{0+}-\tilde{a}_{1}\right)\left(\tilde{a}_{+-}-\tilde{a}_{1}\right)\right)=\\&=e^{\beta\sum_{\ell}\mathcal{E}_{\ell}}\tilde{a}_{1}\left(\tilde{a}_{1}^{2}+\tilde{a}_{-+}\tilde{a}_{+0}+\tilde{a}_{-+}\tilde{a}_{0-}+\tilde{a}_{+0}\tilde{a}_{0-}\right)\\&=e^{\beta\sum_{\ell}\mathcal{E}_{\ell}}\frac{\tilde{a}_{1}}{\Delta_{VMR}}\Delta_{VMR}\left(\left(\Delta_{VMR}\right)^{2}\left(\frac{\tilde{a}_{1}}{\Delta_{VMR}}\right)^{2}+\tilde{a}_{-+}\tilde{a}_{+0}+\tilde{a}_{-+}\tilde{a}_{0-}+\tilde{a}_{+0}\tilde{a}_{0-}\right)\\&=e^{\beta\sum_{\ell}\mathcal{E}_{\ell}}\left(\frac{\tilde{a}_{1}}{\Delta_{VMR}}\right)^{3}\Delta_{VMR}\left(\left(\Delta_{VMR}\right)^{2}+\underset{c_{0}}{\underbrace{\left(\frac{\Delta_{VMR}}{\tilde{a}_{1}}\right)^{2}\tilde{a}_{-+}\tilde{a}_{+0}+\tilde{a}_{-+}\tilde{a}_{0-}+\tilde{a}_{+0}\tilde{a}_{0-}}}\right)
    \end{split}
\end{align}
Since $\tilde{a}_{1}\propto\Delta_{VMR}$ , this expression is well defined for any finite non-zero temperature, with $c_0$ not directly dependent on $\Delta_{VMR}$.

\section{Sufficient condition for oscillations at the high-temperature limit}
The condition for oscillations is $\gamma<0$. Using equation (6) from the main text, while denoting $A_{i}=\int_{\mathcal{E}_{+}}^{\infty}dEe^{-\beta E}a_{i}$, $B_{n\ne \{k,\ell\}}=\int_{\mathcal{E}_{+}}^{\infty}dEe^{-\beta E}b_{k\ell}$, and $\tilde{B}_{n\ne \{k,\ell\}}=\int_{\mathcal{E}_{+}}^{\infty}dEe^{-\beta E}\tilde{b}_{k\ell}$, then the condition $\gamma<0$ becomes:
\footnotesize
\begin{gather}    0>\left(2A_{0}\sum_{i}e^{\beta\mathcal{E}_{i}}+B_{-}\left(e^{\beta\mathcal{E}_{0}}+e^{\beta\mathcal{E}_{+}}\right)+B_{0}\left(e^{\beta\mathcal{E}_{-}}+e^{\beta\mathcal{E}_{+}}\right)+\left(B_{+}+\tilde{B}_{+}\right)\left(e^{\beta\mathcal{E}_{-}}+e^{\beta\mathcal{E}_{0}}\right)\right)^{2}-\nonumber\\-4\left(\prod_{i}e^{\beta\mathcal{E}_{i}}\right)\left(\sum_{j}e^{-\beta\mathcal{E}_{j}}\right)\left(3A_{0}^{2}+2A_{0}\left(B_{-}+B_{0}+B_{+}+\tilde{B}_{+}\right)+A_{1}^{2}+B_{-}B_{0}+\left(B_{-}+B_{0}\right)\left(B_{+}+\tilde{B}_{+}\right)\right)
    \label{eq:osc_cond_beta_no_0}
\end{gather}
\normalsize
In the limit of high temperatures $\beta\to0$ $A_0$ diverges as $\Gamma\left(0,\beta E_{A_{0}}\right)$, with some $E_{A_{0}}>\mathcal{E}_+$, and $\Gamma$ being the incomplete gamma function $\Gamma\left(s,t\right)=\intop_{t}^{\infty}dxe^{-x}x^{s-1}$. This divergence is regulated in the expression of the rates since: $\lim_{\beta\to0}\sqrt{\beta}\Gamma\left(0,\beta E_{A_{0}}\right)=0$. Similarly, when taking the limit in \eqref{eq:osc_cond_beta_no_0}, $A_0$ is multiplied by a term linearly converging to 0. Therefore, when taking $\beta\to0$ the condition for oscillations becomes:
\small
\begin{equation}
    -3A_{1}^{2}+\left(B_{-}-B_{0}\right)^{2}+\left(B_{-}-B_{+}\right)\left(B_{0}-B_{+}\right)+\tilde{B}_{+}\left(\tilde{B}_{+}+\left(B_{+}-B_{0}\right)+\left(B_{+}-B_{-}\right)\right)<0
    \label{eq:sosccond}
\end{equation}
\normalsize
In order to arrive at the expression given in 
(8) in the main text, we look into the expansion of the $b_{k\ell}$ terms. We have for $\mathcal{E}_i<\mathcal{E}_j$:
\begin{equation}
    B_{i}-B_{j}=\frac{4\pi m\nu}{\tilde{Z}_{P}}\intop_{\mathcal{E}_{+}}^{\infty}\frac{2\hbar^{2}\sqrt{\left(E-\mathcal{E}_{+}\right)\left(E-\mathcal{E}_{0}\right)\left(E-\mathcal{E}_{-}\right)}}{\left|D_{T}\right|^{2}}\left(\frac{2\left|u\right|^{2}\hbar^{2}}{2\sqrt{E}}\left(\mathcal{E}_{j}-\mathcal{E}_{i}\right)+\mathcal{O}\left(E^{-\frac{3}{2}}\right)\right)
\end{equation}  
Since for $E>\mathcal{E}_{+}$ we have: $\frac{2\left|u\right|^{2}\hbar^{2}}{2\sqrt{E}}\left(\mathcal{E}_{j}-\mathcal{E}_{i}\right)\le\frac{2\left|u\right|^{2}\hbar^{2}}{2\sqrt{\mathcal{E}_{+}}}\left(\mathcal{E}_{j}-\mathcal{E}_{i}\right)$, then we define:
\begin{align}
    \begin{split}
        \Delta B'_{ij}&=B_{i}-B_{j}+\frac{4\pi m\nu}{\tilde{Z}_{P}}\intop_{\mathcal{E}_{+}}^{\infty}\frac{2\hbar^{2}\sqrt{\left(E-\mathcal{E}_{+}\right)\left(E-\mathcal{E}_{0}\right)\left(E-\mathcal{E}_{-}\right)}}{\left|D_{T}\right|^{2}}\left(\frac{2\left|u\right|^{2}\hbar^{2}}{2\sqrt{\mathcal{E}_{+}}}\left(\mathcal{E}_{j}-\mathcal{E}_{i}\right)-\frac{2\left|u\right|^{2}\hbar^{2}}{2\sqrt{E}}\left(\mathcal{E}_{j}-\mathcal{E}_{i}\right)\right)=\\&=B_{i}-B_{j}+\frac{\left|u\right|^{2}\hbar}{\Im\left(u^{3}\right)}\frac{\mathcal{E}_{j}-\mathcal{E}_{i}}{2\sqrt{2m\mathcal{E}_{+}}}A_{1}-\frac{4\pi m\nu}{\tilde{Z}_{P}}\intop_{\mathcal{E}_{+}}^{\infty}\frac{2\hbar^{2}\sqrt{\left(E-\mathcal{E}_{+}\right)\left(E-\mathcal{E}_{0}\right)\left(E-\mathcal{E}_{-}\right)}}{\left|D_{T}\right|^{2}}\left(\frac{2\left|u\right|^{2}\hbar^{2}}{2\sqrt{E}}\left(\mathcal{E}_{j}-\mathcal{E}_{i}\right)\right)
    \end{split}
\end{align}
s.t. $\Delta B'_{ij}\ge B_{i}-B_{j}$. With this we can write:
\begin{align}
    \begin{split}
&-3A_{1}^{2}+\left(B_{-}-B_{0}\right)^{2}+\left(B_{-}-B_{+}\right)\left(B_{0}-B_{+}\right)+\tilde{B}_{+}\left(\tilde{B}_{+}+\left(B_{+}-B_{0}\right)+\left(B_{+}-B_{-}\right)\right)\le\\\le&-3A_{1}^{2}+\left(\Delta B'_{-0}\right)^{2}+\Delta B'_{-+}\Delta B'_{0+}+\tilde{B}_{+}\left(\tilde{B}_{+}+\left(B_{+}-B_{0}\right)+\left(B_{+}-B_{-}\right)\right)=\\=&-3A_{1}^{2}+A_{1}^{2}\left(\frac{\left|u\right|^{2}\hbar}{\Im\left(u^{3}\right)}\frac{1}{2\sqrt{2m\mathcal{E}_{+}}}\right)^{2}\left(\left(\mathcal{E}_{-}-\mathcal{E}_{0}\right)^{2}+\left(\mathcal{E}_{-}-\mathcal{E}_{+}\right)\left(\mathcal{E}_{0}-\mathcal{E}_{+}\right)+\mathcal{E}_{+}\mathcal{E}_{low-energy}\right)=\\=&-3A_{1}^{2}+3\frac{A_{1}^{2}}{\mathcal{E}_{VDB}}\left(\frac{\left(\mathcal{E}_{-}-\mathcal{E}_{0}\right)^{2}+\left(\mathcal{E}_{-}-\mathcal{E}_{+}\right)\left(\mathcal{E}_{0}-\mathcal{E}_{+}\right)}{\mathcal{E}_{+}}+\mathcal{E}_{low-energy}\right)
    \end{split}
    \label{eq:sht}
\end{align}
where we define:
\begin{gather}
    \mathcal{E}_{low-energy}=\nonumber\\=\frac{\mathcal{E}_{VDB}}{3 A_{1}^{2}}\left(\left(\Delta B'_{-0}\right)^{2}+\Delta B'_{-+}\Delta B'_{0+}+\tilde{B}_{+}\left(\tilde{B}_{+}+\left(B_{+}-B_{0}\right)+\left(B_{+}-B_{-}\right)\right)\right)-\frac{\left(\mathcal{E}_{-}-\mathcal{E}_{0}\right)^{2}+\left(\mathcal{E}_{-}-\mathcal{E}_{+}\right)\left(\mathcal{E}_{0}-\mathcal{E}_{+}\right)}{\mathcal{E}_{+}}
\end{gather}
\normalsize

Notice that if the r.h.s of Eq. \eqref{eq:sht} is negative, then the oscillation condition, Eq. \eqref{eq:sosccond},  automatically holds. Rearranging \eqref{eq:sht} we get equation (8) in the main text. Additionally,  $\mathcal{E}_{VDB},A_{1}^{2}\propto\Delta_{VMR}$, thus making ${\mathcal{E}_{VDB}}/{A_{1}^{2}}$ generally non-zero at points where DB is kept. In order to understand the quantity $\mathcal{E}_{low-energy}$ better, we  plot $\mathcal{E}_{low-energy}/\frac{\left(\mathcal{E}_{-}-\mathcal{E}_{0}\right)^{2}+\left(\mathcal{E}_{-}-\mathcal{E}_{+}\right)\left(\mathcal{E}_{0}-\mathcal{E}_{+}\right)}{\mathcal{E}_{+}}$ (see Figure \ref{fig:E_low_energy}). When this ratio is small, equation (8)  simplifies to:

\begin{gather}
   \mathcal{E}_{VDB}>\frac{\left(\mathcal{E}_--\mathcal{E}_0\right)^2+\left(\mathcal{E}_--\mathcal{E}_+\right)\left(\mathcal{E}_0-\mathcal{E}_+\right)}{\mathcal{E}_+}. \label{eq:bound}
\end{gather}
\begin{figure}[!ht]
    \centering
    \includegraphics[width=0.5\textwidth]{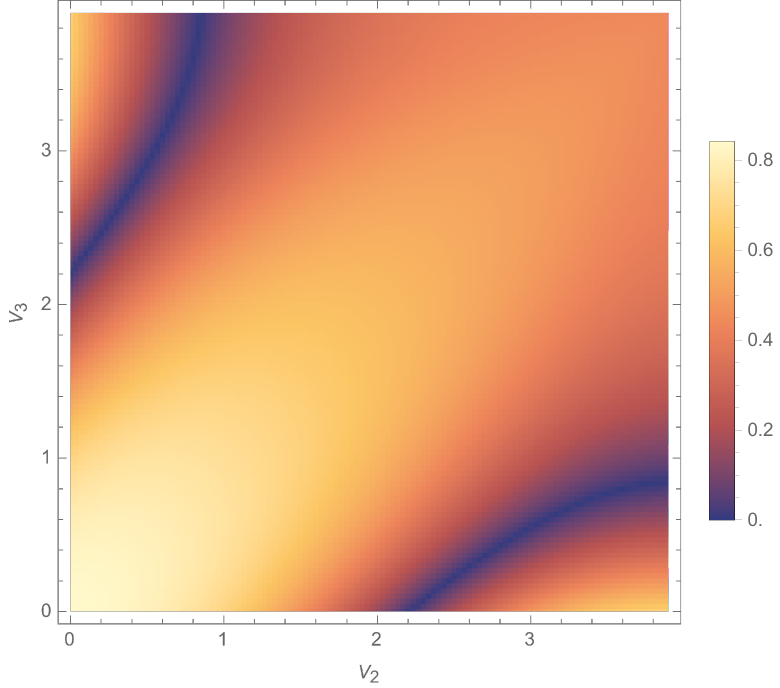}

    \caption{ ${E}_{low- energy}/\frac{\left(\mathcal{E}_--\mathcal{E}_0\right)^2+\left(\mathcal{E}_--\mathcal{E}_+\right)\left(\mathcal{E}_0-\mathcal{E}_+\right)}{\mathcal{E}_+}$ as a function of $V_2$ (x-axis) and $V_3$ (y-axis). 
    Parameters:  $\tau=1.85$, $\phi=0.575$, $V_1=6$. $\hbar=k_B=m=1$.  Max $\Delta \epsilon = \mathcal{E}_+-\mathcal{E}_-$.}
    \label{fig:E_low_energy}
\end{figure}

\newpage
\section{Proof for no oscillations in the low-temperature limit}
For low temperatures, two main results explain the behavior of the rates in the limit of $\beta\to\infty$:
\begin{enumerate}
    \item 
    \begin{equation}
    \label{eqn:rates_low_result1}
    \lim_{\beta\to\infty}\frac{a_{k\ell}}{a_{\ell k}}=\begin{cases}
    \infty & \mathcal{E}_\ell>\mathcal{E}_k\\
0 & \mathcal{E}_\ell<\mathcal{E}_k
\end{cases}
\end{equation}
This means that the transition rates from a higher energetic state to a lower energetic state are much greater than the rates of the reverse process.
\item For $\mathcal{E}_i<\mathcal{E}_j,\mathcal{E}_k<\mathcal{E}_\ell$:
\begin{equation}
    \label{eqn:rates_low_result2}
\frac{B^{L}\left(E_{0}\right)}{B^{U}}\le\lim_{\beta\to\infty}\frac{a_{k\ell}}{a_{ij}}\le\frac{B^{U}}{B^{L}\left(E_{0}\right)}   
\end{equation}
with $0<B^{L}\left(E_{0}\right),B^{U}<\infty$. This means that rates of transition from a higher energy level to a lower one are not significantly bigger or smaller than one another.\\
Combining  both results we find that for $\mathcal{E}_i<\mathcal{E}_j,\mathcal{E}_\ell<\mathcal{E}_k$:
\begin{equation}
    \lim_{\beta\to\infty}\frac{a_{k\ell}}{a_{ij}}=0
\end{equation}
i.e., all transition rates from a low energetic state to a higher one are significantly smaller than any rate from a high energetic state to a lower one.

\end{enumerate}
Using this result we have: $\mathcal{E}_k<\mathcal{E}_\ell$:
\begin{align}
    \begin{split}
        a_{k\ell}^{LT}&\equiv\lim_{\beta\to\infty}\frac{a_{k\ell}}{\omega_{dis}}=\lim_{\beta\to\infty}\frac{a_{k\ell}}{\sum_{i\ne j}a_{ij}}=\lim_{\beta\to\infty}\left(\sum_{\mathcal{E}_{i}<\mathcal{E}_{j}}\frac{a_{ij}}{a_{k\ell}}+\sum_{\mathcal{E}_{i}>\mathcal{E}_{j}}\frac{a_{ij}}{a_{k\ell}}\right)^{-1}=\\&=\left(\sum_{\mathcal{E}_{i}<\mathcal{E}_{j}}\lim_{\beta\to\infty}\frac{a_{ij}}{a_{k\ell}}\right)^{-1}>0
    \end{split}
\end{align}
and similarly $\lim_{\beta\to\infty}\frac{a_{\ell k}}{\omega_{dis}}=0$.
A full proof that $\text{sign}\left(\gamma\left(0\right)\right)=1$ at the limit of $\beta \to \infty$ will follow the proofs of \eqref{eqn:rates_low_result1} and \eqref{eqn:rates_low_result2}.

Before proving 
 \eqref{eqn:rates_low_result1} and \eqref{eqn:rates_low_result2}, we'll see that the rates of transition converge to zero in the limit of $\beta\to\infty$ for this model.
The rates of transition $a_{k\ell}$ are dependent of temperature by factor $\sqrt{\beta}e^{-\beta\left(E-\mathcal{E}_{\ell}\right)}$ where we integrate over $E>\mathcal{E}_{\ell}$. The integral converges for any finite non-zero value of $\beta$, since $\sqrt{\beta}e^{-\beta\left(E-\mathcal{E}_{\ell}\right)}$ is monotonically decreasing for $\beta>\frac{1}{2\left(E-\mathcal{E}_\ell\right)}$, then by dominant convergence theorem we can exchange integration and the limit $\beta\to \infty$. At the limit:
\begin{equation}
    \lim_{\beta\to \infty}\sqrt{\beta}e^{-\beta\left(E-\mathcal{E}_{\ell}\right)}=0
\end{equation}
for any $E>\mathcal{E}_{\ell}$, hence $\lim_{\beta\to \infty}a_{k\ell}=0$

 For the proof of \eqref{eqn:rates_low_result1} and \eqref{eqn:rates_low_result2} we define:
\begin{equation}
    f_{k\ell}\left(E\right)=\frac{1}{\sqrt{E-\max\left\{ \mathcal{E}_{\ell},\mathcal{E}_{k}\right\} }}\frac{\left|T_{k\ell}\left(E\right)\right|^{2}}{\sqrt{\left(E-\mathcal{E}_{\ell}\right)\left(E-\mathcal{E}_{k}\right)}}
\end{equation}
By using the explicit T matrix expressions, we find that for $E\ge \max\left\{ \mathcal{E}_{\ell},\mathcal{E}_{k}\right\}$, this function is bounded, and that $f_{k\ell}>0$.\\
Thus, we'll denote some arbitrary $E_0>\mathcal{E}_{+}$, and define upper and lower bounds for the function:
\begin{align}
    \begin{split}        B_{k\ell}^{L}\left(E_{0}\right)&=\min_{E\in\left[\max\left\{ \mathcal{E}_{\ell},\mathcal{E}_{k}\right\} ,E_{0}\right]}f_{k\ell}\left(E\right)\\B_{k\ell}^{U}&=\max_{E\ge\max\left\{ \mathcal{E}_{\ell},\mathcal{E}_{k}\right\} }f_{k\ell}\left(E\right)
    \end{split}
\end{align}
From that we'll define common bounds:
\begin{align}
    \begin{split}                   B^{L}\left(E_{0}\right)&=\min_{k\ne\ell}B_{k\ell}^{L}\left(E_{0}\right)\\B^{U}&=\max_{k\ne\ell}B_{k\ell}^{U}
    \end{split}
\end{align}
Thus we have $0<B^{L}\left(E_{0}\right)\le B^{U}<\infty$. Defining:
\begin{equation}
    \overline{a}_{k\ell}=\intop_{\max\left\{ \mathcal{E}_{\ell},\mathcal{E}_{k}\right\} }^{\infty}dEe^{-\beta E}\sqrt{E-\max\left\{ \mathcal{E}_{\ell},\mathcal{E}_{k}\right\} }f_{k\ell}\left(E\right)
\end{equation}
gives:
\small
\begin{equation}
    B^{L}\left(E_{0}\right)\intop_{\max\left\{ \mathcal{E}_{\ell},\mathcal{E}_{k}\right\} }^{E_{0}}dEe^{-\beta E}\sqrt{E-\max\left\{ \mathcal{E}_{\ell},\mathcal{E}_{k}\right\} }\le\overline{a}_{k\ell}\le B^{U}\intop_{\max\left\{ \mathcal{E}_{\ell},\mathcal{E}_{k}\right\} }^{\infty}dEe^{-\beta E}\sqrt{E-\max\left\{ \mathcal{E}_{\ell},\mathcal{E}_{k}\right\} }
\end{equation}
\normalsize
changing integration variable $x=E-\max\left\{ \mathcal{E}_{\ell},\mathcal{E}_{k}\right\} $
\begin{equation}
    B^{L}\left(E_{0}\right)e^{-\beta\max\left\{ \mathcal{E}_{\ell},\mathcal{E}_{k}\right\} }\intop_{0}^{E_{0}-\max\left\{ \mathcal{E}_{\ell},\mathcal{E}_{k}\right\} }dEe^{-\beta E}\sqrt{x}\le\overline{a}_{k\ell}\le B^{U}e^{-\beta\max\left\{ \mathcal{E}_{\ell},\mathcal{E}_{k}\right\} }\intop_{0}^{\infty}dxe^{-\beta E}\sqrt{x}
\end{equation}
Now using:
\begin{equation}
    \intop_{0}^{A}dEe^{-\beta E}\sqrt{x}=\frac{\sqrt{\pi}\text{erf}\left(\sqrt{\beta A}\right)-2\sqrt{\beta A}e^{-A\beta}}{2\sqrt{\beta^{3}}}
\end{equation}
with erf being the error function, we get
\small
\begin{equation}
   B^{L}\left(E_{0}\right)e^{-\beta\max\left\{ \mathcal{E}_{\ell},\mathcal{E}_{k}\right\} }\frac{\sqrt{\pi}\text{erf}\left(\sqrt{\beta E_{0}^{k\ell}}\right)-2\sqrt{\beta E_{0}^{k\ell}}e^{-E_{0}^{k\ell}\beta}}{2\sqrt{\beta^{3}}}\le\overline{a}_{k\ell}\le B^{U}e^{-\beta\max\left\{ \mathcal{E}_{\ell},\mathcal{E}_{k}\right\} }\frac{\sqrt{\pi}}{2\sqrt{\beta^{3}}}
\end{equation}
\normalsize
where $E_0^{k\ell}=E_{0}-\max\left\{ \mathcal{E}_{\ell},\mathcal{E}_{k}\right\} >0$.\\
Since $\frac{a_{k\ell}}{a_{ij}}=e^{\beta\left(\mathcal{E}_{\ell}-\mathcal{E}_{j}\right)}\frac{\overline{a}_{k\ell}}{\overline{a}_{ij}}$ :
\begin{equation}
\label{eqn:rates_ratio_general_temp}    e^{\beta\left(\mathcal{E}_{\ell}-\mathcal{E}_{j}\right)}\frac{e^{-\beta\max\left\{ \mathcal{E}_{\ell},\mathcal{E}_{k}\right\} }}{e^{-\beta\max\left\{ \mathcal{E}_{i},\mathcal{E}_{j}\right\} }}\xi_{k\ell}\left(\beta\right)\le\frac{a_{k\ell}}{a_{ij}}\le e^{\beta\left(\mathcal{E}_{\ell}-\mathcal{E}_{j}\right)}\frac{e^{-\beta\max\left\{ \mathcal{E}_{\ell},\mathcal{E}_{k}\right\} }}{e^{-\beta\max\left\{ \mathcal{E}_{i},\mathcal{E}_{j}\right\} }}\left(\xi_{ij}\left(\beta\right)\right)^{-1}
\end{equation}
where 
\begin{equation}
    \xi_{k\ell}\left(\beta\right)\equiv\frac{B^{L}\left(E_{0}\right)}{B^{U}}\frac{\sqrt{\pi}\text{erf}\left(\sqrt{\beta E_{0}^{k\ell}}\right)-2\sqrt{\beta E_{0}^{k\ell}}e^{-E_{0}^{k\ell}\beta}}{\sqrt{\pi}}
\end{equation}
By the properties of the error function, at the limit of low temperatures:
\begin{equation}
    \lim_{\beta\to\infty}\xi_{k\ell}\left(\beta\right)=\frac{B^{L}\left(E_{0}\right)}{B^{U}}
\end{equation}
To finish up the proof, we consider the two cases presented before:
\begin{enumerate}
    \item $i=\ell,j=k$, i.e., considering rates between two states in both directions:
\begin{equation}
\label{eqn:rates_low_result1_pre_limit}
    e^{\beta\left(\mathcal{E}_{\ell}-\mathcal{E}_{k}\right)}\xi_{k\ell}\left(\beta\right)\le\frac{a_{k\ell}}{a_{\ell k}}\le e^{\beta\left(\mathcal{E}_{\ell}-\mathcal{E}_{k}\right)}\left(\xi_{k\ell}\left(\beta\right)\right)^{-1}
\end{equation}   
and at the limit of $\beta\to\infty$ we arrive at \eqref{eqn:rates_low_result1}, with exponential convergence (divergence)
\item $i<j,k<\ell$, i.e., considering rates of transition from a higher energy level to a lower one:
\begin{equation}
    \xi_{k\ell}\left(\beta\right)\le\frac{a_{k\ell}}{a_{ij}}\le\left(\xi_{ij}\left(\beta\right)\right)^{-1}
\end{equation}
which in the limit becomes \eqref{eqn:rates_low_result2}

\end{enumerate}
Finally, we look into the transition rate matrix and note that if we define: $\tilde{M}=\frac{1}{\omega_{dis}}M$ then if $\tilde{M}$ has strictly real eigenvalues then $M$ has only real eigenvalues. Since we have at the limit:
\begin{equation}
    \lim_{\beta\to\infty}\tilde{M}_{ij}=\begin{cases}
a_{ij}^{LT} & i<j\\
0 & i>j
\end{cases}
\end{equation}
then $\tilde{M}$ is triangular. Its eigenvalues are the diagonal elements which are real $\Rightarrow$ $M$ eigenvalues are real $\Rightarrow$ no oscillations in low temperatures.

\section{N - level system toy model}
 A larger number of oscillations during the thermalization time implies a more significant difference between the oscillatory and non-oscillatory decay,  potentially simplifying the possibility of experimentally distinguishing between these two dynamics. Mathematically, the oscillation number is defined as the maximal value of $\left|\frac{\Im\left[\lambda_i\right]}{\Re\left[\lambda_i\right]}\right|$, where $\lambda_i$ is an eigenvalue of $M$. The maximum number of oscillations we show in the main text for the three-level toy model is $\sim 0.16$. As shown below, this number can be increased using the following procedure. 
We consider an N-level system and concentrate on oscillations that involve all its energy levels, that is, $|1\rangle \rightarrow |2\rangle \rightarrow...\rightarrow |N\rangle \rightarrow |1\rangle$. The oscillation strength will be maximized if the oscillation in the opposite direction and oscillations among other levels are forbidden, i.e., $a_{1,2}=a_{2,3}=...=a_{N-1,N}=a_{N,1}=0$. Moreover, for $a_{21}=a_{3,2}=...=a_{N,N-1}=a_{1,N}=a$ the oscillation number is maximized. Notice that this situation corresponds to the strongest violation of detailed balance where $a_{ik}/a_{ki}$ is either zero or infinite. 

In this case, the $M$ matrix is

\begin{equation}
M=a\left(\begin{array}{ccccc}
-1 &1 & 0 &... & 0\\
0 & -1 & 1 & \ddots & 0\\
\vdots & \ddots& \ddots & \ddots & \vdots\\
0 & \ddots & \ddots & -1 & 1\\
1 & 0 & \hdots & 0 & -1
\end{array}\right)  \label{eq:maxei}  
\end{equation}

This is a circulant matrix. The eigenvalues of this matrix are well-known. In particular, the maximal rate between the imaginary and real part of its eigenvalues is $Cot[\frac{\pi}{N}]$. For large $N$, it can be approximated as $N/\pi$ (see figure \ref{fig:maxosc}).   This shows that the oscillation number can be increased to any desired value by increasing the size of the system and maximizing the violation of detailed balance  (see Eq. \eqref{eq:maxei}).

\begin{figure}[!ht]
    \centering
    \includegraphics[width=0.75\textwidth]{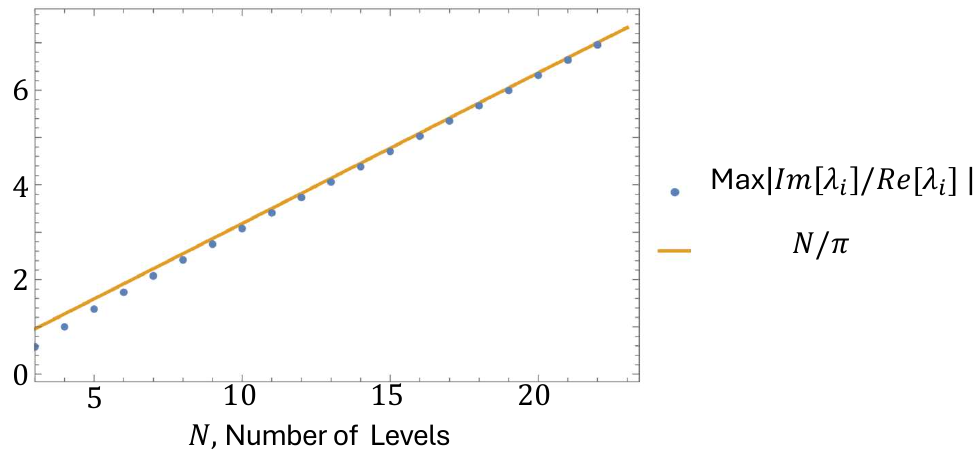}  
    \caption{Maximal oscillation number as a function of the level number for maximum violation of detailed balance.  Eq. \eqref{eq:maxei} shows the form of the transition matrix.}
    \label{fig:maxosc}
\end{figure}

\end{document}